\documentclass{iopart}

\usepackage{graphicx}
\usepackage{latexsym}
\usepackage{amssymb}

\jl{6}        
\eqnobysec    

\def\beq{\begin{equation}}
\def\eeq{\end{equation}}
\def\rmd{{\rm d}}

\begin{document}

\title[Quadrupolar bodies in a Kerr spacetime]
{Extended bodies in a Kerr spacetime: exploring the role of a general quadrupole tensor}

\author{
Donato Bini$^{1,2}$ and
Andrea Geralico${}^{3,2}$
}

\address{${}^1$\
Istituto per le Applicazioni del Calcolo ``M. Picone,'' CNR, 
I--00185 Rome, Italy}

\address{
  ${}^2$\
  ICRA,
  ``Sapienza'' University of Rome  I--00185 Rome, Italy
}

\address{
  $^3$
  Physics Department,
  ``Sapienza'' University of Rome, I--00185 Rome, Italy}

\ead{binid@icra.it}

\begin{abstract}
The equatorial motion of extended bodies in a Kerr spacetime is investigated in the framework of the Mathisson-Papapetrou-Dixon model, including the full set of effective components of the quadrupole tensor.
The numerical integration of the associated equations shows the specific role of the mass and current quadrupole moment components.
While most of the literature on this topic is limited to spin-induced (purely electric) quadrupole tensor, the present analysis highlights the effect of a completely general quadrupole tensor on the dynamics.
The contribution of the magnetic-type components is indeed related to a number of interesting features, e.g., enhanced inward/outward spiraling behavior of the orbit and spin-flip-like effects, which may have observational counterparts.  
Finally, the validity limit of the Mathisson-Papapetrou-Dixon model is also discussed through explicit examples.
\end{abstract}

\pacno{04.20.Cv}

\section{Introduction}

Quadrupolar effects on the motion of extended bodies can be very important in many realistic astrophysical situations, e.g., when an extended body is moving around a central source, like ordinary or neutron stars and binary pulsar systems orbiting the Galactic Center black hole (Sgr A$^*$) \cite{falcke,lyne}.
For instance, there is a special interest in studying the orbits of the compact objects close to Sgr A$^*$ \cite{muno} because of the increasing accuracy in sub-milli-arcsecond astrometry by the near-infrared detectors \cite{eise} and the improved potentiality of the next-generation radiotelescopes, e.g., the Square Kilometer Array (SKA) \cite{ska}.

A general relativistic model describing the motion of extended bodies endowed with multipolar structure is due to Dixon \cite{dixon64,dixon69,dixon70,dixon73,dixon74}, who generalized to higher multipole moments the pioneering works of Mathisson \cite{math37}, Papapetrou \cite{papa51,corpapa51}, Pirani \cite{pir56} and Tulczyjew \cite{tulc59} for spinning particles. 
The model equations have been treated by different approximation schemes both analytically and numerically in astrophysically relevant background spacetimes. 
Examples of numerical studies of the full nonlinear equations in the case of purely spinning bodies can be found, e.g., in Refs. \cite{maeda,sem99,ver,hartl1,hartl2}.

In this paper we study the dynamics of an extended body endowed with both spin and quadrupole moment in a Kerr spacetime according to the Mathisson-Papapetrou-Dixon (MPD) model. 
The body is assumed to be \lq\lq quasi-rigid,'' i.e., with constant components of the quadrupole tensor with respect to the frame adapted to the body's generalized 4-momentum, according to the definition of Ehlers and Rudolph \cite{ehlers77}.
As it is well known, the quadrupole tensor has in general 20 independent components and shares the same algebraic symmetries of the Riemann tensor. 
However, it enters the MPD equations only through certain contractions with the Riemann tensor and its covariant derivative, so that of its original 20 independent components only those obeying also the symmetries of the MPD equations will survive. 
We have shown in Ref. \cite{quadrupkerr1} that the number of relevant components is actually reduced to 10 in vacuum, as expected from the standard post-Newtonian formulation of motion of many-body systems, leading to the definition of an \lq\lq effective'' quadrupole tensor.
Furthermore, we assume the motion to be confined on the equatorial plane and the spin vector of the body to be aligned with the axis of rotation of the central object. 
The number of nonvanishing effective components of the quadrupole tensor then reduces from 10 to 5: 3 belong to the mass quadrupole moment, and 2 to the current quadrupole moment.
We have explored in Ref. \cite{quadrupkerr1} both analytically and numerically the case in which the quadrupole tensor is completely specified by two independent components only, both of them belonging to the mass quadrupole part.
In this paper we complete our previous analysis by considering all 5 nonzero components and studying their effect on the dynamics of the body through numerical simulations. 
We are also motivated in widening the discussion to the case of a general quadrupole tensor by related works concerning the inclusion in the post-Newtonian dynamics of a two-body system of the quadrupole corrections induced by the spin (see, e.g., Refs. \cite{porto06,porto06prl,faye06,djs,porto08,steinhoff08,hergt08,steinhoff09,hinderer}).

We find a number of interesting features. The body deviates from geodesic behavior due to its structure, as expected, but deviations can become more and more enhanced for special quadrupole configurations. This fact has the consequence that the evolution may drive the body out of the possibility to be described within the MPD model, since, at a certain time, the spin length becomes no longer negligible with respect to the natural length scale associated with the gravitational field. In this case one must abandon the description of the body in the framework of the MPD model and use instead the theory of first order gravitational perturbations to take into due account the backreaction effects on the background metric. In previously examined solutions \cite{quadrupschw,quadrupkerr1} concerning bodies endowed with special quadrupolar configurations we generally found oscillations about the reference geodesic motion, which we chose to be spatially circular. 
This fact supported the rough idea that the spin and the quadrupole of the body reflect themselves into a sort of \lq\lq uncertainty'' of the position close to the geodesic path. The present analysis confirms but also enlarges this picture.
In fact, in general, when the conditions for the MPD model are still satisfied, the orbit of the body spirals around the black hole either approaching its outer horizon or escaping. 
Finally, we recall that an analytical approach to this problem is also possible.
However, the main formulas obtained in Ref. \cite{quadrupkerr1} for the associated perturbative solution are general enough to simply incorporate the present case, so we refer to it for such an analysis, which is beyond the aim of this work.

\section{MPD model for quadrupolar bodies}

Consider an extended body endowed with structure up to the quadrupole, following the description due to Dixon.
In the quadrupole approximation, Dixon's equations are
\begin{eqnarray}
\label{papcoreqs1}
\frac{{\rm D}P^{\mu}}{\rmd \tau} & = &
- \frac12 \, R^\mu{}_{\nu \alpha \beta} \, U^\nu \, S^{\alpha \beta}
-\frac16 \, \, J^{\alpha \beta \gamma \delta} \, \nabla^\mu R_{\alpha \beta \gamma \delta}
\nonumber\\
& \equiv & F^\mu_{\rm (spin)} + F^\mu_{\rm (quad)} \,,
\\
\label{papcoreqs2}
\frac{{\rm D}S^{\mu\nu}}{\rmd \tau} & = & 
2 \, P^{[\mu}U^{\nu]}+
\frac43 \, J^{\alpha \beta \gamma [\mu}R^{\nu]}{}_{\gamma \alpha \beta}
\nonumber\\
&\equiv & D^{\mu \nu}_{\rm (spin)} + D^{\mu \nu}_{\rm (quad)} \,,
\end{eqnarray}
where $P^{\mu}=m u^\mu$ (with $u \cdot u = -1$) is the total 4-momentum of the body with mass $m$, $S^{\mu \nu}$ is a (antisymmetric) spin tensor, $J^{\alpha\beta\gamma\delta}$ is the quadrupole tensor, and $U^\mu=\rmd z^\mu/\rmd\tau$ is the timelike unit tangent vector of the \lq\lq center of mass line'' (with parametric equations $x^\mu=z^\mu(\tau)$) used to make the multipole reduction, parametrized by the proper time $\tau$. 

In order the model to be mathematically self-consistent the following additional conditions should be imposed \cite{tulc59,dixon64} (see also Ref. \cite{semerak} for a comparative discussion of different choices of supplementary conditions adopted in the literature)
\beq
\label{tulczconds}
S^{\mu\nu}u{}_\nu=0\,.
\eeq
Consequently, the spin tensor can be fully represented by a spatial vector (with respect to $u$),
\beq
S(u)^\alpha=\frac12 \eta(u)^\alpha{}_{\beta\gamma}S^{\beta\gamma}=[{}^{*_{(u)}}S]^\alpha\,,\qquad
\eta(u)_{\alpha\beta\gamma}=\eta_{\mu\alpha\beta\gamma}u^\mu\,,
\eeq
where $\eta(u)_{\alpha\beta\gamma}$ is the spatial (with respect to $u$) unit volume 3-form with $\eta_{\alpha\beta\gamma\delta}=\sqrt{-g} \epsilon_{\alpha\beta\gamma\delta}$ the unit volume 4-form and $\epsilon_{\alpha\beta\gamma\delta}$ ($\epsilon_{0123}=1$) the Levi-Civita alternating symbol. As standard, hereafter we denote the spacetime dual of a tensor (built up with $\eta_{\alpha\beta\gamma\delta}$) by a $^*$, whereas the 
spatial dual of a spatial tensor with respect to $u$ (built up with $\eta(u)_{\alpha\beta\gamma}$) by $^{*_{(u)}}$.
It is also useful to introduce the signed magnitude $s$ of the spin vector
\beq
\label{sinv}
s^2=S(u)^\beta S(u)_\beta = \frac12 S_{\mu\nu}S^{\mu\nu}\,, 
\eeq
which is in general not constant along the trajectory of the extended body. 

The quadrupole tensor $J^{\alpha\beta\gamma\delta}$ has the same algebraic symmetries as the Riemann tensor, i.e.,
\beq
J^{\alpha\beta\gamma\delta}=J^{[\alpha\beta][\gamma\delta]}=J^{\gamma\delta\alpha\beta}\,,\qquad
J^{[\alpha\beta\gamma]\delta}=0\,,
\eeq
leading to 20 independent components.
However, it has been shown in Ref. \cite{quadrupkerr1} that, since $J$ enters the MPD equations only through certain combinations (contractions), the number of effective components actually reduces from 20 to 10.
Therefore, it is worth to use directly the following \lq\lq effective'' quadrupole tensor (still denoted by $J$) which also shares all the symmetries underlying the MPD equations
\begin{eqnarray}
\label{deco_bar_u5}
J^{\alpha\beta}{}_{\gamma\delta}&=&4\bar u^{[\alpha}[X(\bar u)]^{\rm STF}{}^{\beta]}{}_{[\gamma}\bar u_{\delta]}
+2\bar u^{[\alpha}[W(\bar u)]^{\rm STF}{}^{\beta]}{}_\sigma \eta(\bar u)^{\sigma}{}_{ \gamma\delta}\nonumber\\
&&
+2\bar u_{[\gamma}[W(\bar u)]^{\rm STF}{}_{\delta]}{}_\sigma \eta(\bar u)^{\sigma \alpha\beta}\,,
\end{eqnarray} 
where $X(\bar u)$ and $W(\bar u)$ are symmetric and trace-free (STF) spatial tensors as measured by an observer with $4$-velocity $\bar u$.
When the observer $\bar u=u$, i.e., when the observer is at rest with respect to the body, the spatial tensors $X(\bar u)$ and $W(\bar u)$ have an intrinsic meaning, representing the mass quadrupole moment and the flow (or current) quadrupole moment (see, e.g., Ref. \cite{ehlers77}), and reduce to the corresponding Newtonian quantities in the Newtonian limit.
Therefore, we will assume $\bar u=u$ hereafter.

We also recall the standard $1+3$ representation of the Riemann tensor in vacuum, i.e.,
\begin{eqnarray}
\label{rieman_deco}
R^{\alpha\beta\gamma\delta}&=&-\eta(u)^{\alpha \beta \mu}\eta(u)^{\gamma\delta\nu}E(u)_{\mu\nu}
+2u^{[\alpha}H(u)^{\beta]}{}_\sigma \eta(u)^{\sigma \gamma\delta}\nonumber\\
&&+2u^{[\gamma}H(u)^{\delta]}{}_\sigma \eta(u)^{\sigma \alpha\beta}
-4u^{[\alpha}E(u)^{\beta][\gamma}u^{\delta]}\,,
\end{eqnarray}
in terms of its electric ($E(u)$) and magnetic ($H(u)$) parts defined by
\beq
E(u)_{\alpha\beta}=R_{\alpha\mu\beta\nu}u^\mu u^\nu\,,\qquad
H(u)_{\alpha\beta}=-[R^*]_{\alpha\mu\beta\nu}u^\mu u^\nu\,.
\eeq

Finally, when the background spacetime has Killing vectors, there are conserved quantities along the motion \cite{ehlers77}. For example, in the case of stationary axisymmetric spacetimes with coordinates adapted to the spacetime symmetries,  $\xi=\partial_t$ is the timelike Killing vector and $\eta=\partial_\phi$ is the azimuthal Killing vector. The corresponding conserved quantities are the total energy $E$ and the angular momentum $J$, namely
\begin{eqnarray}
\label{totalenergy}
E&=&-\xi_\alpha P^\alpha +\frac12 S^{\alpha\beta}F^{(t)}_{\alpha\beta}\,,\qquad
F^{(t)}_{\alpha\beta}=\nabla_\beta \xi_\alpha=g_{t[\alpha,\beta]}\,,\nonumber\\
J&=&\eta_\alpha P^\alpha -\frac12 S^{\alpha\beta}F^{(\phi)}_{\alpha\beta}\,,\qquad
F^{(\phi)}_{\alpha\beta}=\nabla_\beta \eta_\alpha=g_{\phi[\alpha,\beta]}\,, 
\end{eqnarray}
where $F^{(t)}$ and $F^{(\phi)}$ are the Papapetrou fields associated with the Killing vectors.

\section{Dynamics of extended bodies in the equatorial plane of a Kerr spacetime}

The Kerr metric  in standard Boyer-Lindquist coordinates is given by
\begin{eqnarray}
\rmd s^2 &=& -\left(1-\frac{2Mr}{\Sigma}\right)\rmd t^2 -\frac{4aMr}{\Sigma}\sin^2\theta\rmd t\rmd\phi+ \frac{\Sigma}{\Delta}\rmd r^2\nonumber\\
&& 
+\Sigma\rmd \theta^2+\frac{(r^2+a^2)^2-\Delta a^2\sin^2\theta}{\Sigma}\sin^2 \theta \rmd \phi^2\,,
\end{eqnarray}
where $\Delta=r^2-2Mr+a^2$ and $\Sigma=r^2+a^2\cos^2\theta$; here $a$ and $M$ are the specific angular momentum and total mass of the spacetime solution. The event horizon and inner horizon are located at $r_\pm=M\pm\sqrt{M^2-a^2}$.

Let us introduce the zero angular momentum observer (ZAMO) family of fiducial observers, with 4-velocity
\beq
\label{n}
n=N^{-1}(\partial_t-N^{\phi}\partial_\phi)\,,
\eeq
where $N=(-g^{tt})^{-1/2}$ and $N^{\phi}=g_{t\phi}/g_{\phi\phi}$ are the lapse and shift functions, respectively. A suitable orthonormal frame adapted to ZAMOs is given by
\beq
\label{ZAMO-frame}
e_{\hat t}=n\,, \quad
e_{\hat r}=\frac1{\sqrt{g_{rr}}}\partial_r\,,\quad
e_{\hat \theta}=\frac1{\sqrt{g_{\theta \theta }}}\partial_\theta\,, \quad
e_{\hat \phi}=\frac1{\sqrt{g_{\phi \phi }}}\partial_\phi\,,
\eeq
with dual 
\beq\fl\qquad
\label{ZAMO-frame-dual}
\omega^{{\hat t}}=N\rmd t\,, \quad 
\omega^{{\hat r}}=\sqrt{g_{rr}}\rmd r\,, \quad
\omega^{{\hat \theta}}= \sqrt{g_{\theta \theta }} \rmd \theta\,, \quad
\omega^{{\hat \phi}}=\sqrt{g_{\phi \phi }}(\rmd \phi+N^{\phi}\rmd t)\,.
\eeq

The ZAMOs are accelerated with acceleration $a(n)=\nabla_n n$ and locally non-rotating, in the sense that their vorticity vector $\omega(n)^\alpha$ vanishes, but they have a nonzero expansion tensor $\theta(n)_{\alpha\beta}$; the latter, in turn, can be completely described by an expansion vector $\theta_{\hat \phi}(n)^\alpha=\theta(n)^\alpha{}_\beta\,{e_{\hat\phi}}^\beta$, that is
\beq
\label{exp_zamo}
\theta(n) = e_{\hat\phi}\otimes\theta_{\hat\phi}(n)
+\theta_{\hat\phi}(n)\otimes e_{\hat\phi}
\,.
\eeq 
The trace of the expansion tensor $\theta(n)^\alpha{}_\alpha$ turns out to be zero.

The nonzero ZAMO kinematical quantities (i.e., acceleration and expansion) all belong to the $r$-$\theta$ 2-plane of the tangent space \cite{mfg,idcf1,idcf2,bjdf}, i.e.,
\begin{eqnarray}
\label{accexp}
a(n) & = & a(n)^{\hat r} e_{\hat r} + a(n)^{\hat\theta} e_{\hat\theta}
\equiv\partial_{\hat r}(\ln N) e_{\hat r} + \partial_{\hat\theta}(\ln N)  e_{\hat\theta}
\,,
\nonumber\\
\theta_{\hat\phi}(n) & = & \theta_{\hat\phi}(n)^{\hat r}e_{\hat r} + \theta_{\hat\phi}(n)^{\hat\theta}e_{\hat \theta} 
\equiv-\frac{\sqrt{g_{\phi\phi}}}{2N}\,(\partial_{\hat r} N^\phi e_{\hat r} + \partial_{\hat\theta} N^\phi e_{\hat \theta})
\,.
\end{eqnarray}
In the static limit (as it is the case of a Schwarzschild field) $N^\phi\to0$ and the expansion vector $\theta_{\hat\phi}(n)$ identically vanishes. 

It is convenient to introduce the Lie relative curvature for orbits on the equatorial plane \cite{idcf1,idcf2}, which is purely radial with value
\beq
k_{\rm (lie)}=-\partial_{\hat r} \ln \sqrt{g_{\phi\phi}}\,.
\eeq
Finally, timelike circular geodesics $U_{\pm}$ have unit tangent vector
\beq
U_{\pm}=\gamma_{\pm} [e_{\hat t} +\nu_{\pm} e_{\hat \phi}]\,, \qquad
\gamma_\pm=(1-\nu_{\pm}^2)^{-1/2}\,,
\eeq
where $\nu_{\pm}$ denote the linear velocities associated with co-rotating $(+)$ and counter-rotating $(-)$ orbits
\beq
\nu_\pm =\frac{a^2\mp2a\sqrt{Mr}+r^2}{\sqrt{\Delta}(a\pm r\sqrt{r/M})}\,.
\eeq 
In the static case $\nu_\pm\to\pm\nu_K$, with $\nu_K=\sqrt{M/(r-2M)}$.

The ZAMO kinematical quantities as well as the nonvanishing frame components of the Riemann tensor are listed in Appendix A.

\subsection{Orbit of the extended body}

Let the world line of the extended body with unit tangent vector $U$ be confined on the equatorial plane, i.e.,  
\beq
\label{polarnu}
U=\gamma(U,n) [n+ \nu(U,n)]\,,
\eeq
with
\beq
\label{polarnu2}
\nu(U,n)\equiv \nu^{\hat r}e_{\hat r}+\nu^{\hat \phi}e_{\hat \phi} 
=  \nu (\cos \alpha e_{\hat r}+ \sin \alpha e_{\hat \phi})\,,
\eeq
where $\gamma(U,n)=1/\sqrt{1-||\nu(U,n)||^2}\equiv\gamma$ is the Lorentz factor and the abbreviated notation $\nu^{\hat a}\equiv\nu(U,n)^{\hat a}$ has been used.  Similarly $\nu \equiv ||\nu(U,n)||$ and $\alpha$ are the magnitude of the spatial velocity $\nu(U,n)$ and its
polar angle  measured clockwise from the positive $\phi$ direction in the $r$-$\phi$ tangent plane respectively, while $\hat\nu\equiv\hat \nu(U,n)$ is the associated unit vector.
Note that $\alpha=\pi/2$ corresponds to azimuthal motion with respect to the ZAMOs, while 
$\alpha=0,\pi$ correspond to (outward/inward) radial motion with respect to the ZAMOs.

A convenient adapted frame to $U$ is given by
\beq\fl\qquad
\label{U_frame}
E_1\equiv \hat \nu ^\perp=\sin\alpha  e_{\hat r}- \cos\alpha  e_{\hat \phi}\,,\quad
E_2=\gamma [\nu n+\hat \nu]\,,\quad
E_3=-e_{\hat \theta}\,.
\eeq
A similar decomposition holds for the 4-momentum $P=mu$ for equatorial motion, i.e.,  
\beq
\label{polarnuu}
u=\gamma_u [n +\nu_u\hat \nu_u]\,, \qquad
\gamma_u=(1-\nu_u^2)^{-1/2}\,,
\eeq
with
\beq
\label{polarnuu2}
\hat \nu (u,n)\equiv \hat \nu_u=\cos\alpha_u e_{\hat r}+ \sin\alpha_u e_{\hat \phi}\,.
\eeq
An orthonormal frame adapted to $u\equiv e_0$ is then built with the spatial triad
\beq\fl\qquad
\label{uframe}
e_1\equiv \hat \nu_u^\perp=\sin\alpha_u e_{\hat r}- \cos\alpha_u e_{\hat \phi}\,,\quad
e_2=\gamma_u [\nu_u n +\hat \nu_u]\,,\quad
e_3=-e_{\hat \theta}\,.
\eeq
The dual frame of $\{e_\alpha\}$ will be denoted by $\{\omega^\alpha\}$, with $\omega^0=-u^\flat$, $n^\flat$ being the fully covariant representation of $n$.

The projection of the spin tensor into the local rest space of $u$ defines the spin vector $S(u)$ (hereafter simply denoted by $S$, for short).
Therefore, when decomposed with respect to the frame (\ref{uframe}) adapted to $u$, it writes as
\beq
S=S^1e_1+S^2e_2+S^3e_3\,.
\eeq

\subsection{Setting the body's spin and quadrupole}

In the following we will consider the special case in which the spin vector is aligned along the spacetime rotation axis, i.e.,
\beq
S=se_3\,.
\eeq
When decomposed with respect to the frame adapted to $u$, the spin terms defined in Eqs. (\ref{papcoreqs1}) and (\ref{papcoreqs2}) are thus given by
\beq
\label{fspinframeu}
F_{\rm (spin)}=F_{\rm (spin)}^0u+F_{\rm (spin)}^1e_1+F_{\rm (spin)}^2e_2\,,
\eeq
and 
\beq
\label{dspinframeu}
D_{\rm (spin)}=-\omega^0\wedge{\mathcal E}(u)_{\rm (spin)}\,,
\eeq
with
\beq
\label{dspinframeu2}
{\mathcal E}(u)_{\rm (spin)}={\mathcal E}(u)_{\rm (spin)}{}_1\omega^{1}+{\mathcal E}(u)_{\rm (spin)}{}_2\omega^{2}\,,
\eeq
respectively.
The explicit expressions for the components are listed in Appendix B. 

Furthermore, we will assume the quadrupole tensor given by Eq. (\ref{deco_bar_u5}) with $\bar u=u$ having constant frame components with respect to the frame (\ref{uframe}) adapted to $u$ as the most natural and simplifying choice. 
According to the terminology introduced in Ref. \cite{ehlers77}, the extended body should be termed in this case as \lq\lq quasi-rigid.'' 
In order that the motion be confined on the equatorial plane we must require $X(u)_{13}=X(u)_{23}=0$ and $W(u)_{11}=W(u)_{22}=W(u)_{12}=0$.
The quadrupole force and torque with respect to the frame adapted to $u$ are thus given by
\beq
\label{fquadframeu}
F_{\rm (quad)}=F_{\rm (quad)}^0u+F_{\rm (quad)}^1e_1+F_{\rm (quad)}^2e_2\,,
\eeq
and
\beq
\label{dquadframeu}
D_{\rm (quad)}=-\omega^0\wedge{\mathcal E}(u)_{\rm (quad)}+{}^{*_{(u)}}{\mathcal B}(u)_{\rm (quad)}\,,
\eeq
with
\begin{eqnarray}
\label{dquadframeu2}
{\mathcal E}(u)_{\rm (quad)}&=&{\mathcal E}(u)_{\rm (quad)}{}_1 \omega^1 +{\mathcal E}(u)_{\rm (quad)}{}_2 \omega^2\,,\nonumber\\
{\mathcal B}(u)_{\rm (quad)}&=&{\mathcal B}(u)_{\rm (quad)}{}_3 \omega^3\,,
\end{eqnarray}
so that ${\mathcal E}(u)_{\rm (quad)}\cdot {\mathcal B}(u)_{\rm (quad)}=0$, respectively.
The explicit expressions for the components of the quadrupole force and torque are listed in Appendix B.

\section{Numerical integration of the full set of MPD equations}

Under the assumptions discussed in the previous section on the structure of the body as well as on its motion the whole set of MPD equations (\ref{papcoreqs1})--(\ref{tulczconds}) reduces to 
\begin{eqnarray}\fl\quad 
\label{setfin}
\frac{\rmd m}{\rmd \tau} &=& 
F_{\rm (spin)}^0 + F_{\rm (quad)}^0\,,\nonumber\\
\fl\quad
\frac{\rmd \alpha_u}{\rmd \tau} &=& 
-\frac{\gamma}{\nu_u}\left[\nu\cos(\alpha_u+\alpha)-\nu_u\right]\theta_{\hat\phi}(n)^{\hat r}
+\frac{\gamma}{\nu_u}\left(\sin\alpha_ua(n)^{\hat r}+\nu\nu_u\sin\alpha k_{\rm (lie)}\right)\nonumber\\
\fl\quad
&&
-\frac{1}{m\gamma_u\nu_u}(F_{\rm (spin)}^1 + F_{\rm (quad)}^1)\,,\nonumber\\
\fl\quad
\frac{\rmd \nu_u}{\rmd \tau} &=& 
-\frac{\gamma}{\gamma_u^2}\left[\cos\alpha_ua(n)^{\hat r}+\nu\sin(\alpha_u+\alpha)\theta_{\hat\phi}(n)^{\hat r}\right]
+\frac{1}{m\gamma_u^2}(F_{\rm (spin)}^2 + F_{\rm (quad)}^2)\,,\nonumber\\
\fl\quad
\frac{\rmd s}{\rmd \tau} &=&
{\mathcal B}(u)_{\rm (quad)}^3\,,
\end{eqnarray}
together with the following two compatibility conditions coming from the spin evolution equations 
\begin{eqnarray} 
\label{compatib}
0&=&m({\mathcal E}(u)_{\rm (spin)}^1+{\mathcal E}(u)_{\rm (quad)}^1)+s(F_{\rm (spin)}^2 + F_{\rm (quad)}^2)\,,\nonumber\\
0&=&-m({\mathcal E}(u)_{\rm (spin)}^2+{\mathcal E}(u)_{\rm (quad)}^2)+s(F_{\rm (spin)}^1 + F_{\rm (quad)}^1)\,.
\end{eqnarray}
The evolution equation for the spin invariant implies that the quadrupolar structure of the body is responsible for the onset of spin angular momentum, if the body is initially not spinning.

Equations (\ref{compatib}) give two algebraic relations involving the remaining unknowns $\nu$ and $\alpha$.
After some manipulation we find
\beq
\label{nuandalpha}
\tan\alpha=\frac{A+B\gamma}{C+D\gamma}\,, \qquad
\gamma=\frac{k_1+\sqrt{k_1^2+k_2k_3}}{k_2}\,,
\eeq
where
\beq 
k_1=AB+CD\,,\quad
k_2=1-B^2-D^2\,,\quad
k_3=1+A^2+C^2\,,
\eeq
and 
\begin{eqnarray}\fl\quad 
\label{ABC_etc}
\lambda A&=&
\left(sF_{\rm (quad)}^1-m{\mathcal E}(u)_{\rm (quad)}^2\right)
\left\{
[m^2-s^2E_{\hat r\hat r}+s^2\gamma_u^2(E_{\hat r\hat r}-E_{\hat \theta\hat \theta})]\sin\alpha_u\right.\nonumber\\
\fl\quad
&&\left.
-s^2H_{\hat r\hat \theta}\gamma_u^2\nu_u(1+\sin^2\alpha_u)
\right\}\nonumber\\
\fl\quad
&&
+\left(sF_{\rm (quad)}^2+m{\mathcal E}(u)_{\rm (quad)}^1\right)
[m^2-s^2E_{\hat \theta\hat \theta}-s^2H_{\hat r\hat \theta}\nu_u\sin\alpha_u]\gamma_u\cos\alpha_u
\,,\nonumber\\
\fl\quad
B&=&\frac{(m^2-s^2E_{\hat r\hat r})\nu_u\sin\alpha_u+s^2H_{\hat r\hat \theta}}{m^2-s^2E_{\hat \theta\hat \theta}-s^2H_{\hat r\hat \theta}\nu_u\sin\alpha_u}\,,\nonumber\\
\fl\quad
\lambda C&=&
\left(sF_{\rm (quad)}^1-m{\mathcal E}(u)_{\rm (quad)}^2\right)\cos\alpha_u
\left\{
m^2-s^2E_{\hat \theta\hat \theta}\right.\nonumber\\
\fl\quad
&&\left.
-s^2\gamma_u^2\nu_u[(E_{\hat r\hat r}+2E_{\hat \theta\hat \theta})\nu_u+H_{\hat r\hat \theta}\sin\alpha_u]
\right\}\nonumber\\
\fl\quad
&&
-\left(sF_{\rm (quad)}^2+m{\mathcal E}(u)_{\rm (quad)}^1\right)
[m^2-s^2E_{\hat \theta\hat \theta}-s^2H_{\hat r\hat \theta}\nu_u\sin\alpha_u]\gamma_u\sin\alpha_u
\,,\nonumber\\
\fl\quad
D&=&\frac{m^2+s^2(E_{\hat r\hat r}+E_{\hat \theta\hat \theta})}{m^2-s^2E_{\hat \theta\hat \theta}-s^2H_{\hat r\hat \theta}\nu_u\sin\alpha_u}\nu_u\cos\alpha_u\,,\nonumber\\
\fl\quad
\lambda &=&\gamma_u[m^2-s^2E_{\hat \theta\hat \theta}-s^2H_{\hat r\hat \theta}\nu_u\sin\alpha_u]\left\{
m^2+\frac12s^2E_{\hat \theta\hat \theta}(1-3\gamma_u^2)\right.\nonumber\\
\fl\quad
&&\left.
-s^2\gamma_u^2\nu_u\left[\frac12\nu_u\cos2\alpha_u(2E_{\hat r\hat r}+E_{\hat \theta\hat \theta})+2H_{\hat r\hat \theta}\sin\alpha_u\right]
\right\}\,.
\end{eqnarray}

Note that the following relations hold
\beq
\nu\sin\alpha=\frac{A}{\gamma}+B\,,\qquad
\nu\cos\alpha=\frac{C}{\gamma}+D\,,
\eeq
and the first equation of Eqs. (\ref{setfin}) governing the mass evolution can be written
\beq
\label{massevol}
\frac{\rmd m}{\rmd \tau} = c_1A+c_2C+ F_{\rm (quad)}^0\,,
\eeq
with
\begin{eqnarray}
c_1&=&-s^2\gamma_u\nu_u\cos\alpha_u[E_{\hat r\hat r}+2E_{\hat \theta\hat \theta}+H_{\hat r\hat \theta}\nu_u\sin\alpha_u]\,,\nonumber\\
c_2&=&-s^2\gamma_u[(E_{\hat r\hat r}-E_{\hat \theta\hat \theta})\nu_u\sin\alpha_u
-H_{\hat r\hat \theta}(1+\nu_u^2\sin^2\alpha_u)]\,,
\end{eqnarray}
once the dependence on $\nu$ and $\alpha$ in the component $F_{\rm (spin)}^0$ has been eliminated.
The rhs of Eq. (\ref{massevol}) vanishes for vanishing quadrupole (i.e., $F_{\rm (quad)}^a=0={\mathcal E}(u)_{\rm (quad)}^a$, so that $A=0=C$), yielding the well known constant mass result for a purely spinning particle.

Finally, in order to perform a numerical integration of Eqs. (\ref{setfin}) the evolution equations $U=\rmd x^\alpha/\rmd\tau$ must be also taken into account, i.e., 
\beq\fl\quad
\label{Ucompts}
\frac{\rmd t}{\rmd \tau} =\frac{\gamma}{N}\,,\qquad
\frac{\rmd r}{\rmd \tau} =\frac{\gamma\nu\cos\alpha}{\sqrt{g_{rr}}}\,,\qquad
\frac{\rmd \phi}{\rmd \tau}= \frac{\gamma}{\sqrt{g_{\phi\phi}}}\left(\nu\sin\alpha-\frac{\sqrt{g_{\phi\phi}}N^\phi}{N}\right)\,.
\eeq

As a consistency check, the total energy $E$ and angular momentum $J$ given by Eq. (\ref{totalenergy}), i.e., 
\begin{eqnarray}
\label{EandJ}
E&=&N\gamma_u\left[m+s(\nu_u\sin\alpha_u a(n)^{\hat r}+\theta_{\hat\phi}(n)^{\hat r})\right]-N^\phi J
\,,\nonumber\\
J&=&\gamma_u\sqrt{g_{\phi\phi}}\left[m\nu_u\sin\alpha_u-s(k_{\rm (lie)}+\nu_u\sin\alpha_u\theta_{\hat\phi}(n)^{\hat r})\right]
\,,
\end{eqnarray}
remain constant and equal to their initial values.
Solving the above relations for $\gamma_u$ and $\nu_u\sin\alpha_u$ leads to
\begin{eqnarray}
\qquad\gamma_u&=&\frac{(m-s\theta_{\hat\phi}(n)^{\hat r})(E+N^\phi J)\sqrt{g_{\phi\phi}}-sNa(n)^{\hat r}J}{\sqrt{g_{\phi\phi}}N\{m^2+s^2[a(n)^{\hat r}k_{\rm (lie)}-(\theta_{\hat\phi}(n)^{\hat r})^2]\}}
\,,\nonumber\\
\nu_u\sin\alpha_u&=&\frac{sk_{\rm (lie)}\sqrt{g_{\phi\phi}}(E+N^\phi J)+NJ(m+s\theta_{\hat\phi}(n)^{\hat r})}{(m-s\theta_{\hat\phi}(n)^{\hat r})(E+N^\phi J)\sqrt{g_{\phi\phi}}-sNa(n)^{\hat r}J}
\,,
\end{eqnarray}
which allow to express in terms of $E$ and $J$ the frame components with respect to ZAMOs of the generalized 4-momentum $P=mu$ of the particle, as form Eqs. (\ref{polarnuu}) and (\ref{polarnuu2}).
In absence of quadrupole (implying constant mass and spin magnitude) this is enough to obtain from the radial component an effective potential associated with radial motion \cite{tod,maeda2,steinhoff2012} (even if, to be really useful, the latter should refer to $U$ and not to $P$).
In the presence of the quadrupole, instead, both $m$ and $s$ in general vary with $r$, so that a similar study cannot be performed. 
Nevertheless, a simple inspection of the above energy and angular momentum conservation laws allows one to get some general features of particle's motion.
For simplicity, in the Schwarzschild case Eqs. (\ref{EandJ}) reduce to 
\beq
\label{EandJsch}
\fl\qquad
E=m\gamma_u\left[N+\frac{s}{m} (r\nu_u\sin\alpha_u) \frac{M}{r^3}\right] 
\,,\qquad
J=m\gamma_u\left[r\nu_u\sin\alpha_u+N\frac{s}{m}\right]
\,,
\eeq
so that their (constant) ratio turns out to depend on the signed spin length $s/m$ instead of $m$ and $s$ separately
\beq
\frac{J}{E}=\frac{r\nu_u\sin\alpha_u+N\frac{s}{m}}{N+\frac{s}{m} (r\nu_u\sin\alpha_u) \frac{M}{r^3}}\,.
\eeq
For large value of $r$ we have $N\to1$ and $r\nu_u\sin\alpha_u\sim r$, whence  
\beq
\frac{J}{E}\sim r+\frac{s}{m}\,,
\eeq
implying that $s/m$ has to become increasingly negative in this limit.
For $r$ approaching the horizon, instead, $N\to0$ and
\beq
\frac{J}{E}\sim \left[1+\frac{s}{m}\frac{M}{r^3}\right]^{-1}\,,
\eeq
implying that $s/m$ has to be positive, attaining a constant limiting value.

\section{Numerical analysis and validity of the MPD model}

In this section we discuss the main features of motion of extended bodies endowed with both spin and quadrupole moment on the equatorial plane of a Kerr spacetime by numerically integrating the full system of MPD equations for selected values of the parameters representing the structure of the body.
Initial conditions are fixed so that the orbit is (initially) tangent to a circular geodesic at a given radius. 
As a result of the evolution, the motion is in general spiraling either inward or outward depending on the chosen set of quadrupole values. 
In order to check the validity of the assumptions underlying the MPD model, we complement each plot of the orbit with the corresponding evolution of the length scale associated with the spin, which should maintain small with respect to the characteristic length of the background.
As an indicator we have fixed a threshold for $s/m$ of some percent of $M$ (typically between $1\%$ and $2\%$).

\begin{figure} 
\typeout{*** EPS figure 1}
\begin{center}
$\begin{array}{cc}
\includegraphics[scale=0.3]{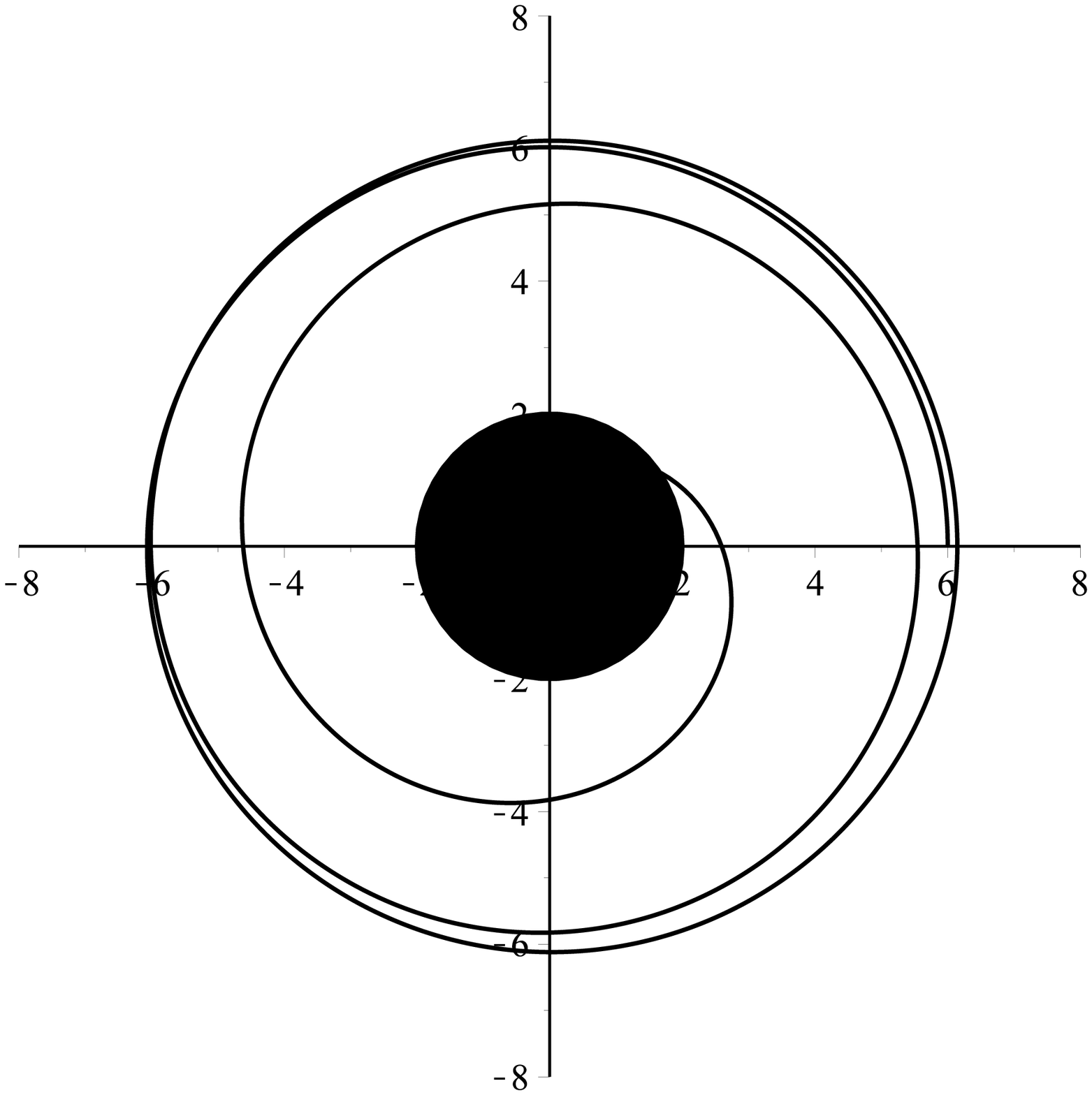}&\quad
\includegraphics[scale=0.3]{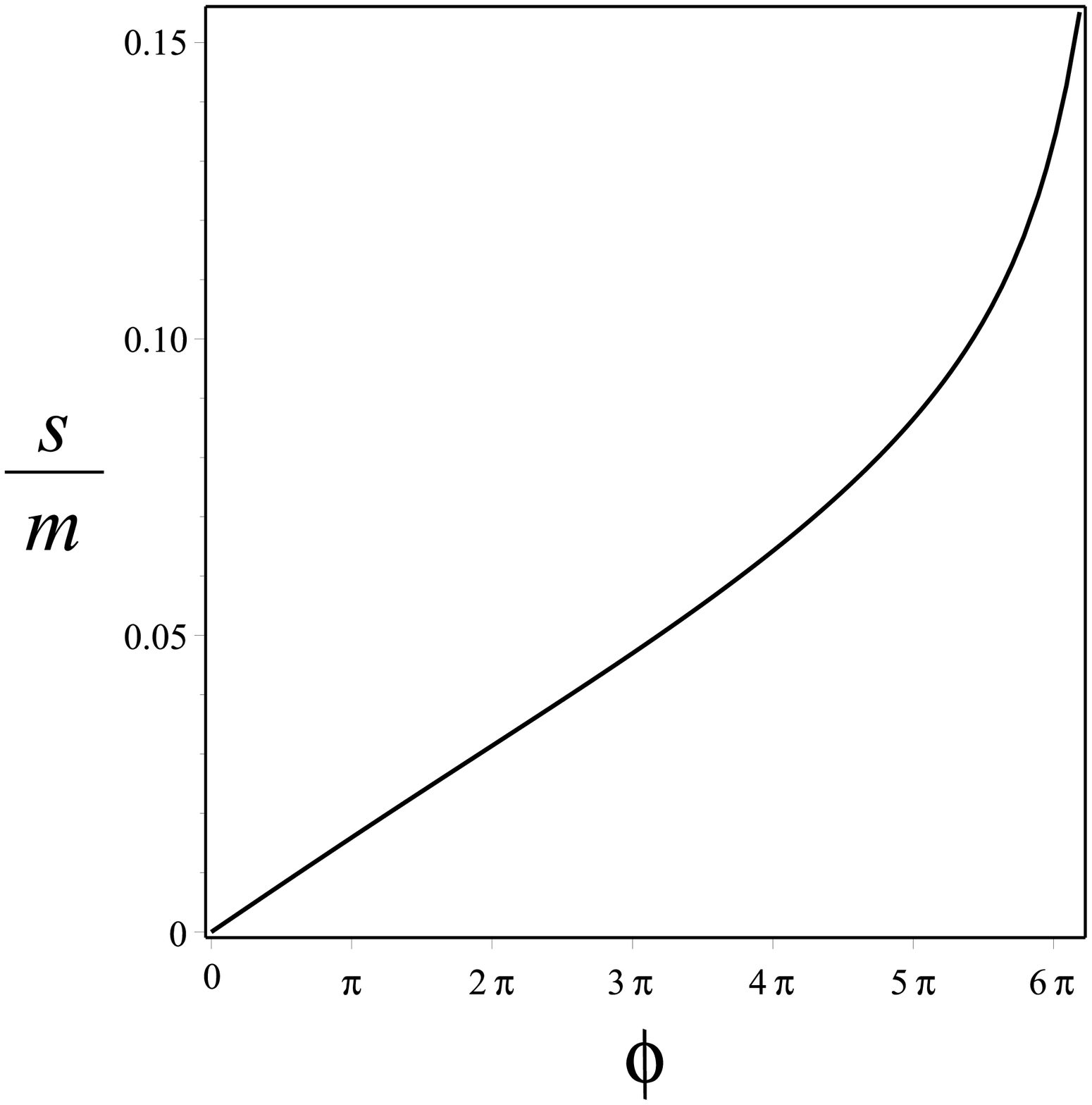}\\[.4cm]
\quad\mbox{(a)}\quad &\quad \mbox{(b)}
\end{array}$\\
\end{center}
\caption{The orbit of the extended body is shown in the Schwarzschild case in panel (a) for the following choice of parameters and initial conditions: $r_0/M=6$, $X(u)_{11}=X(u)_{12}=X(u)_{22}=W(u)_{13}=W(u)_{23}=0.01m_0M^2$, and $r(0)=r_0$, $\phi(0)=0$, $\alpha_u(0)=\pi/2$, $\nu_u(0)=\nu_K=0.5$, $m(0)=m_0$, $s(0)=10^{-6}$ (in units of $m_0M$).
Panel (b) shows instead the corresponding behavior of the length scale associated with the spin, i.e, $s/m$.
The initial value of $s$ has been taken exaggerately small to enhance the effect of the quadrupole.
}
\label{fig:1}
\end{figure}

\begin{figure} 
\typeout{*** EPS figure 2}
\begin{center}
$\begin{array}{cc}
\includegraphics[scale=0.3]{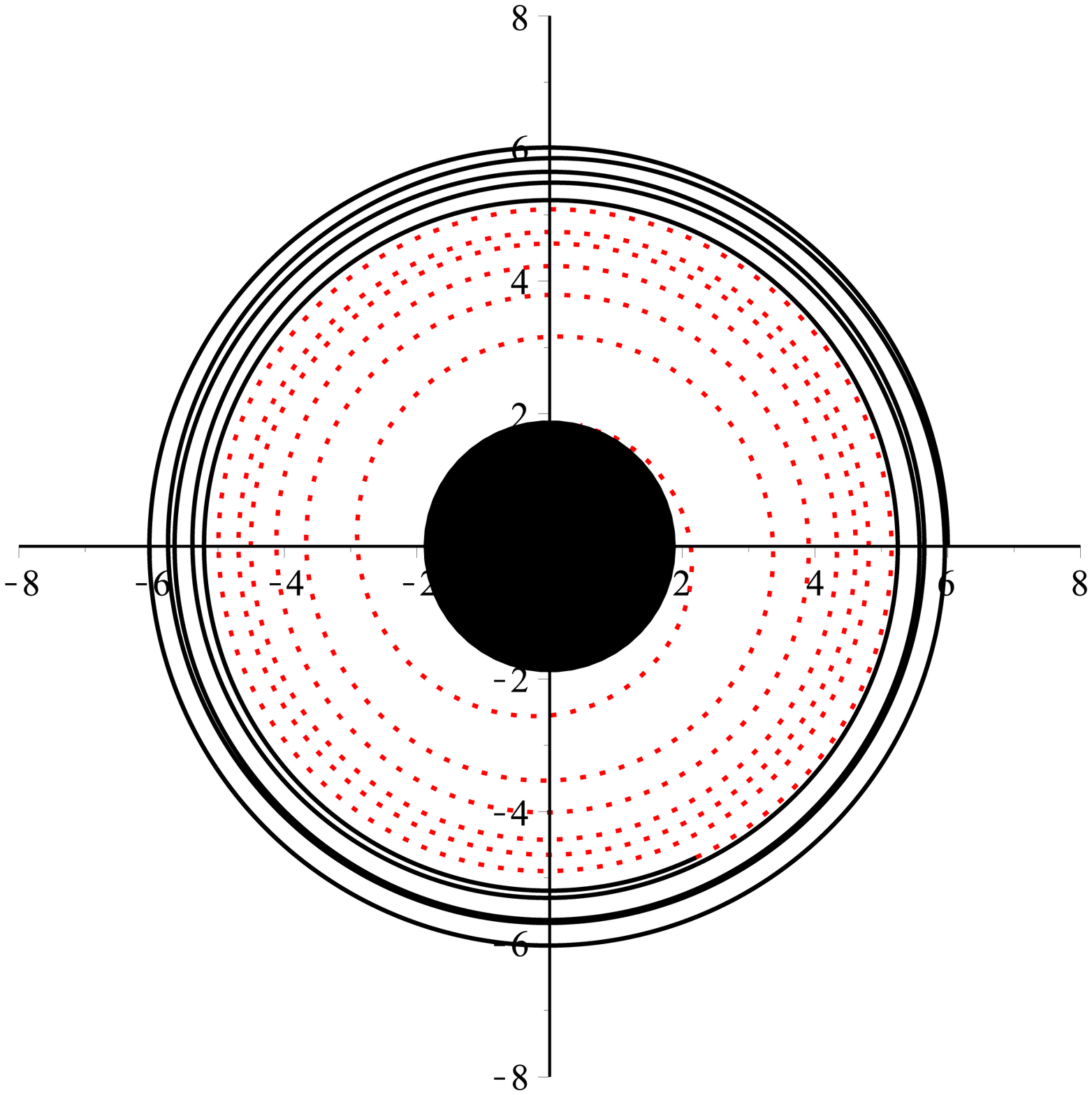}&\quad
\includegraphics[scale=0.3]{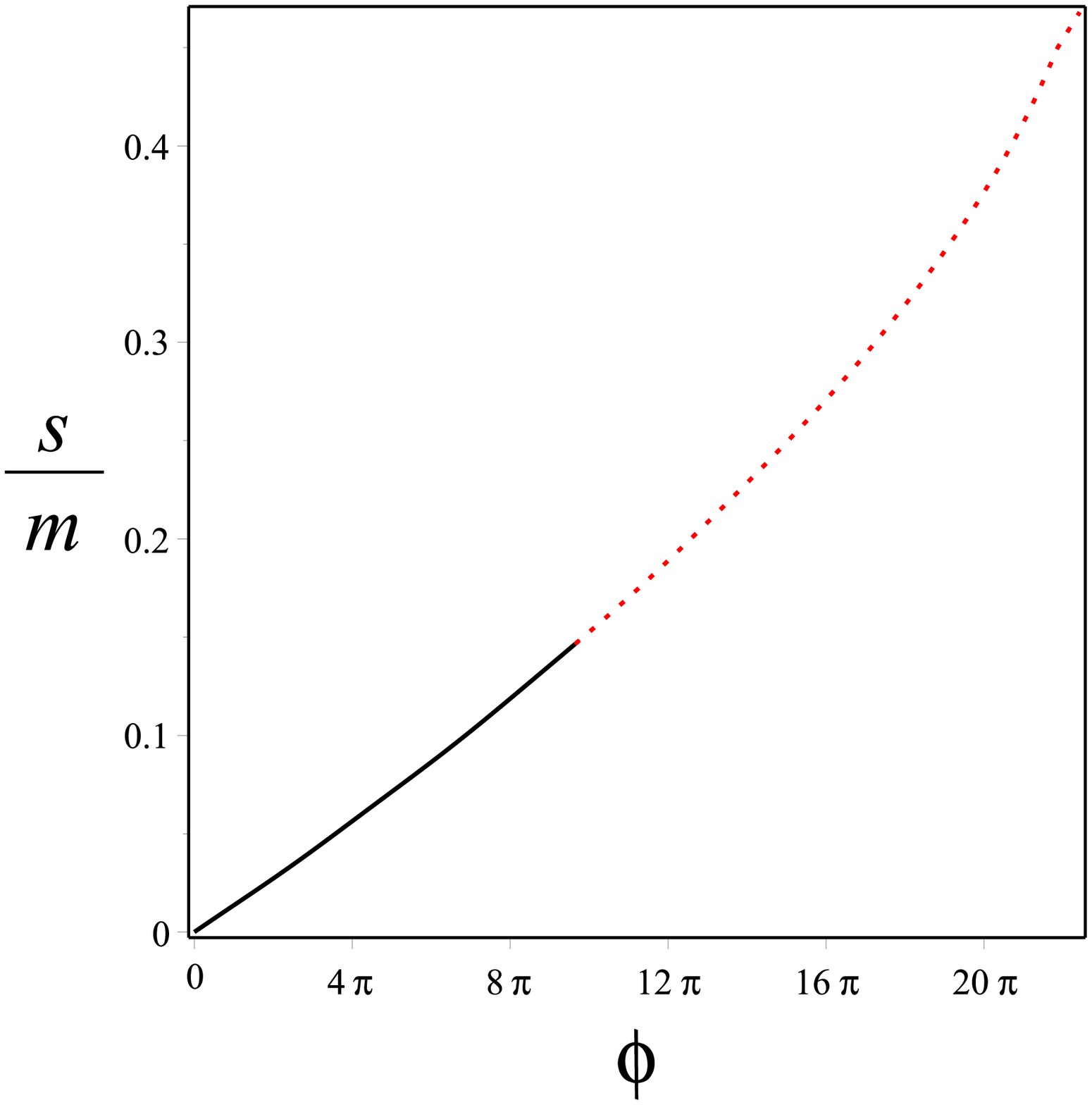}\\[.4cm]
\quad\mbox{(a)}\quad &\quad \mbox{(b)}
\end{array}$\\
\end{center}
\caption{The orbit of the extended body is shown in the Kerr case in panel (a) for the following choice of parameters and initial conditions: $a/M=0.5$, $r_0/M=6$, $X(u)_{11}=X(u)_{12}=X(u)_{22}=W(u)_{13}=W(u)_{23}=0.01m_0M^2$, and $r(0)=r_0$, $\phi(0)=0$, $\alpha_u(0)=\pi/2$, $\nu_u(0)=\nu_+\approx0.45165$, $m(0)=m_0$, $s(0)=10^{-6}$ (in units of $m_0M$).
Panel (b) shows instead the corresponding behavior of the length scale associated with the spin.
Solid curves correspond to the region of validity of the MPD model as $s/(mM)\lesssim0.15$.
Such curves have been analytically continued (dotted curves) also in a situation in which the MPD model cannot be applied.
}
\label{fig:2}
\end{figure}

We begin by considering the Schwarzschild case. In order to enhance the contribution due to the quadrupole we set a very small initial value for the spin magnitude $s$, whereas the five components of the quadrupole tensor are assumed to be all nonzero and with the same value. We find that the orbit spirals inward up to the horizon with $s$ remaining small enough, thus preserving the limit of validity of the MPD model (see Fig. \ref{fig:1}). 
The same spiraling behavior is exhibited also in the Kerr case (see Fig. \ref{fig:2}), but the spin magnitude increases more rapidly because of the spin-spin interaction, so that the MPD model becomes no more valid after few revolutions. 
Nevertheless, we have continued the orbit (dotted curves) to visualize the region wherein backreaction effects cannot be neglected.

\begin{figure} 
\typeout{*** EPS figure 3}
\begin{center}
$\begin{array}{cc}
\includegraphics[scale=0.3]{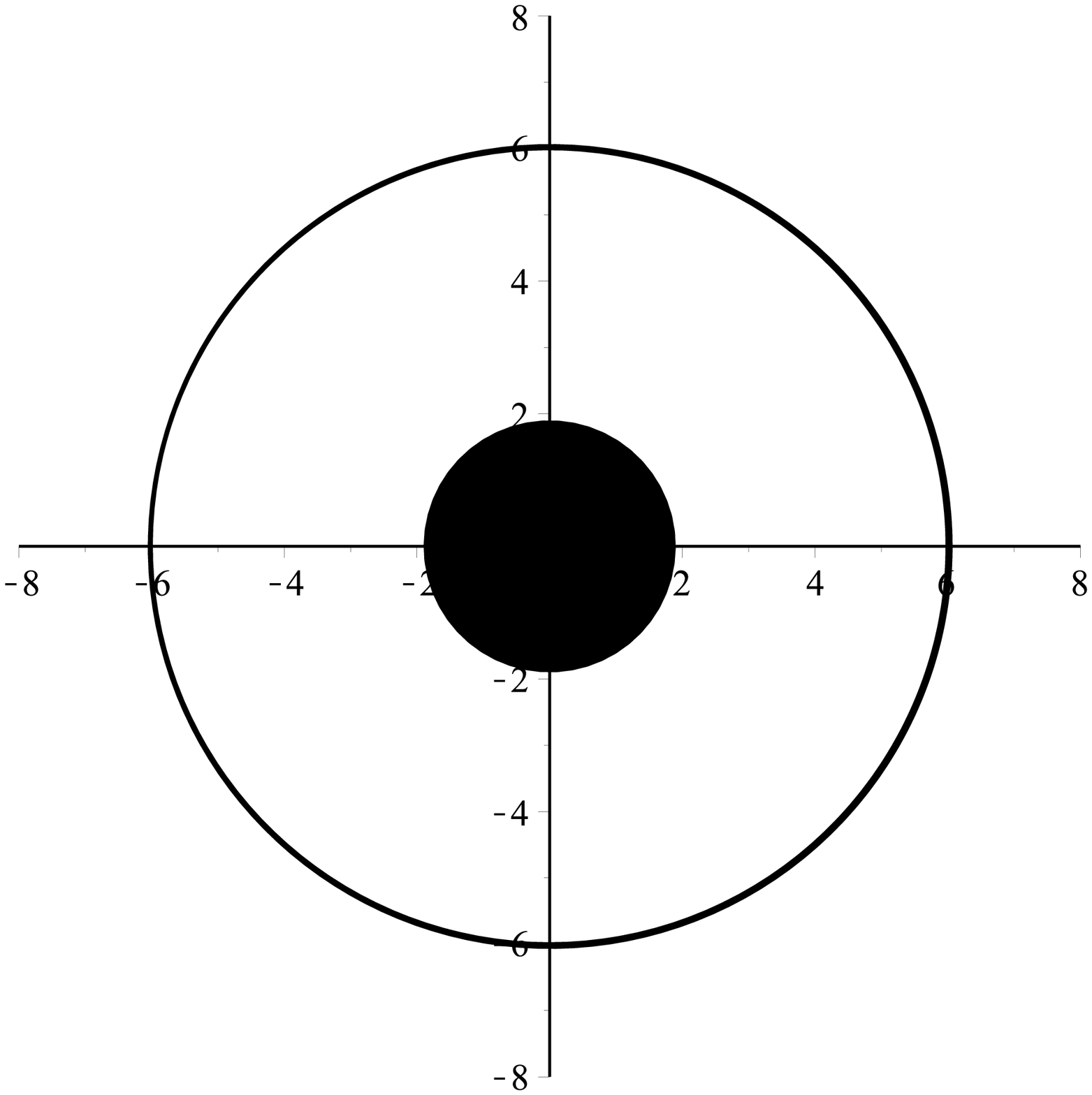}&\quad
\includegraphics[scale=0.3]{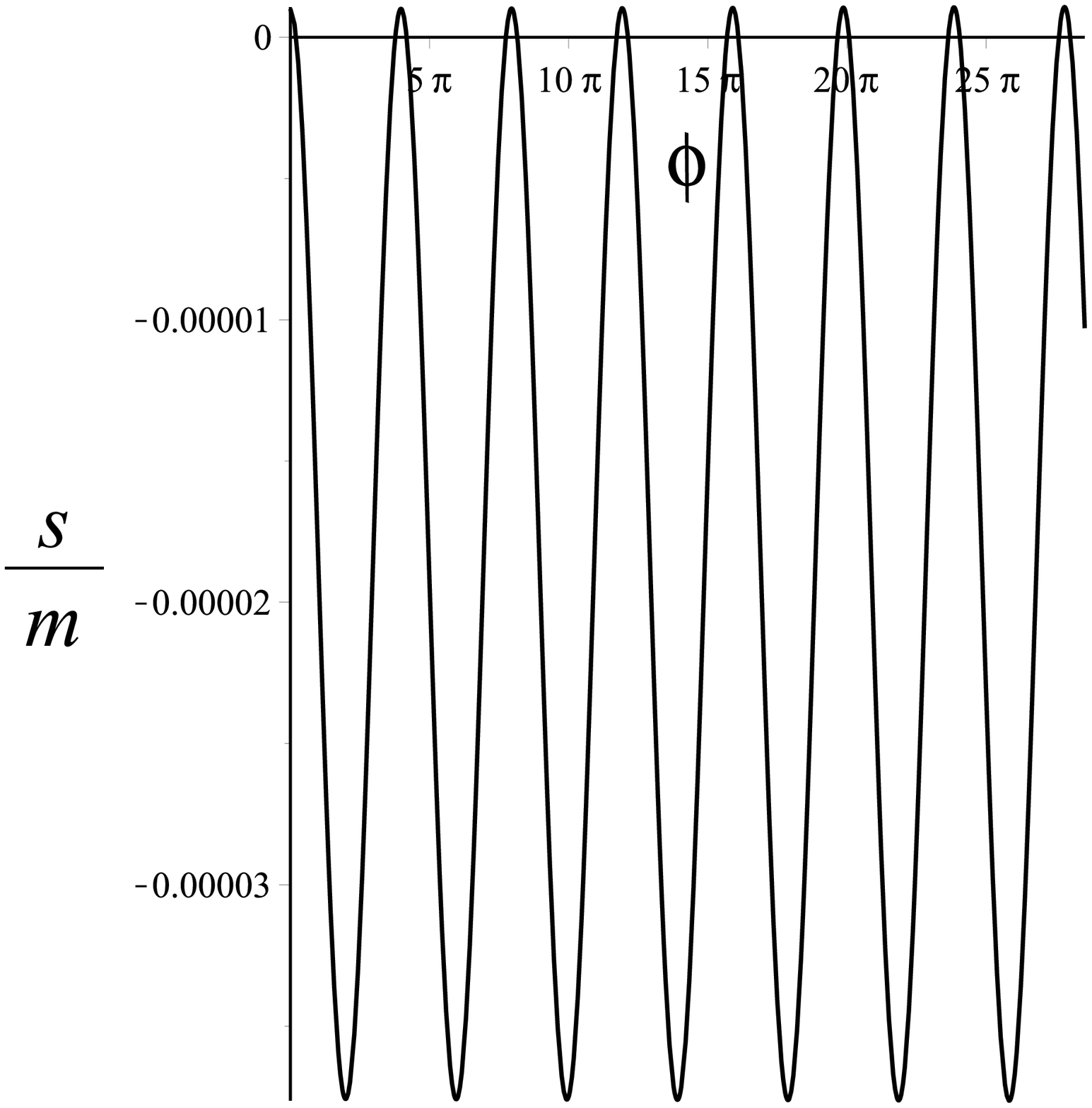}\\[.4cm]
\quad\mbox{(a)}\quad &\quad \mbox{(b)}
\end{array}$\\
\end{center}
\caption{The orbit of the extended body is shown in the Kerr case in panel (a) for the same choice of parameters and initial conditions as in Fig. \ref{fig:2}, except for the values of the quadrupole tensor components, which are now specified by $X(u)_{11}=-X(u)_{22}=0.01m_0M^2$ (the remaining components being set equal to zero, i.e., $X(u)_{12}=W(u)_{13}=W(u)_{23}=0$).
The actual orbit oscillates around the reference circular geodesic with oscillation amplitude of about $4\%$ of the initial radius.
Panel (b) shows instead the corresponding behavior of the length scale associated with the spin.
}
\label{fig:3}
\end{figure}

\begin{figure} 
\typeout{*** EPS figure 4}
\begin{center}
$\begin{array}{cc}
\includegraphics[scale=0.3]{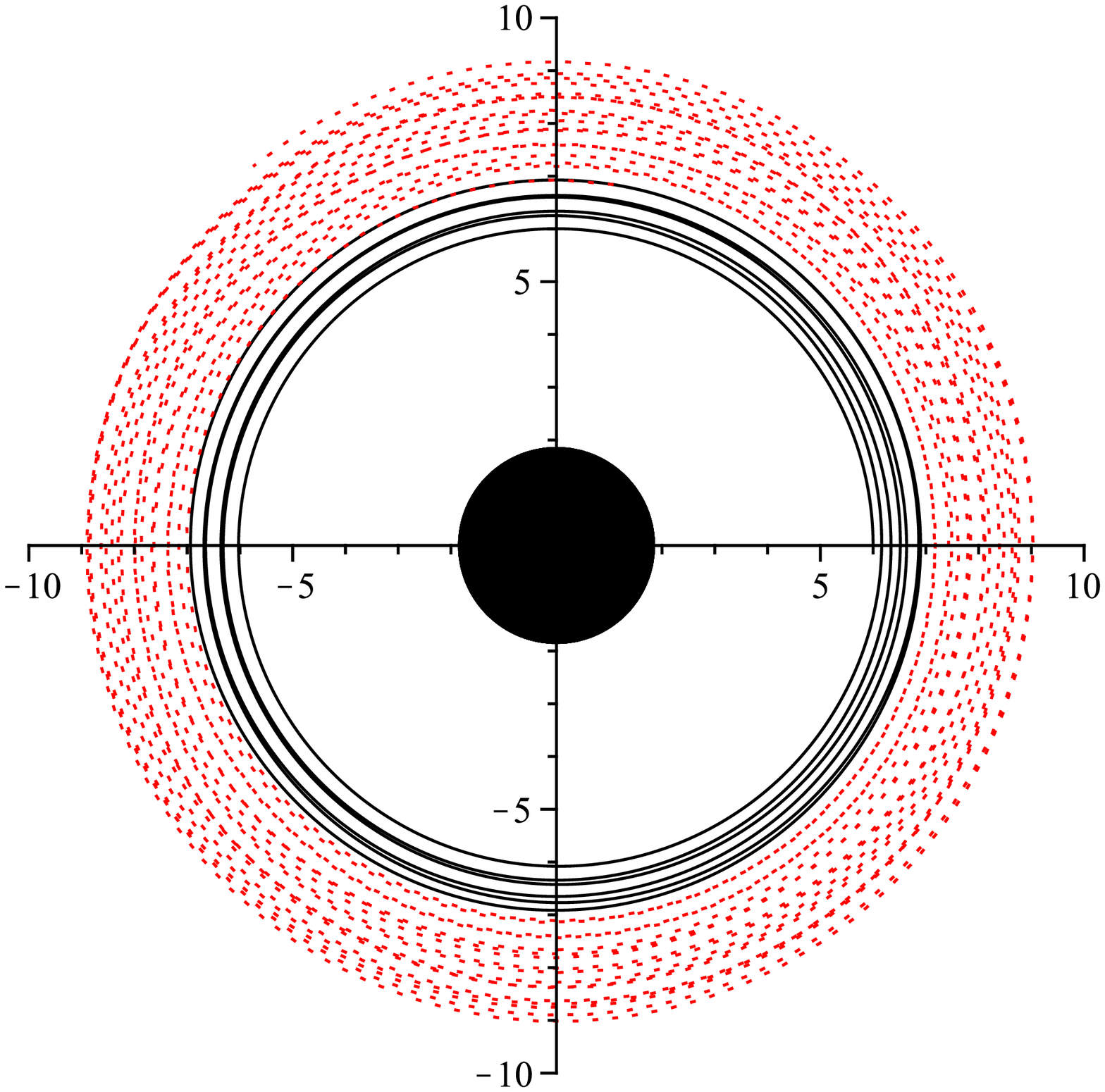}&\quad
\includegraphics[scale=0.3]{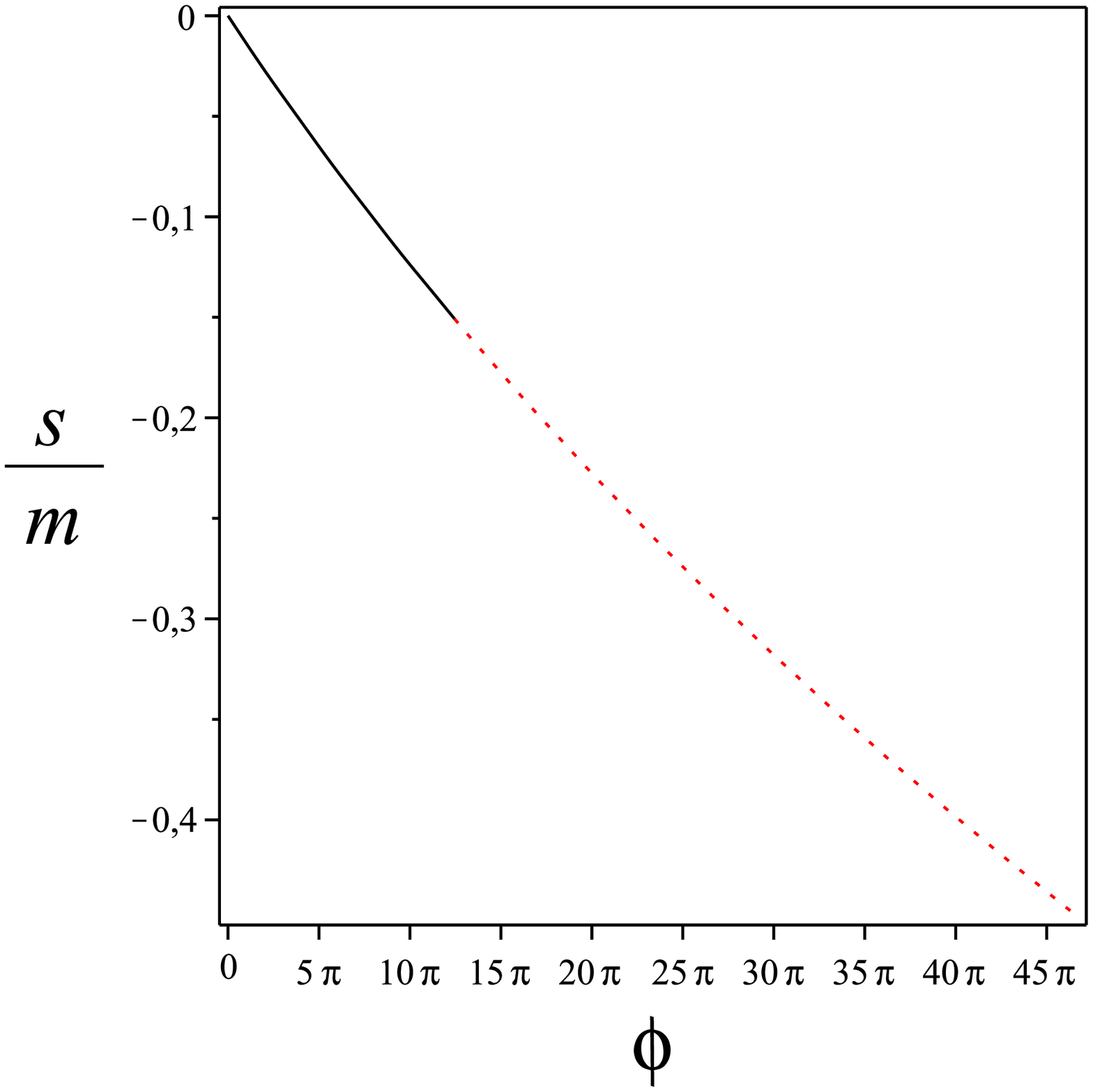}\\[.4cm]
\quad\mbox{(a)}\quad &\quad \mbox{(b)}
\end{array}$\\
\end{center}
\caption{The orbit of the extended body is shown in the Kerr case in panel (a) for the same choice of parameters and initial conditions as in Fig. \ref{fig:2}, except for the values of the quadrupole tensor components, which are now specified by $X(u)_{11}=-X(u)_{12}=X(u)_{22}=-W(u)_{13}=-W(u)_{23}=0.01m_0M^2$.
Panel (b) shows instead the corresponding behavior of the length scale associated with the spin.
Reversing the sign of the non-diagonal components of the quadrupole tensor thus implies that the body spirals outward.
}
\label{fig:4}
\end{figure}

\begin{figure} 
\typeout{*** EPS figure 5}
\begin{center}
$\begin{array}{cc}
\includegraphics[scale=0.3]{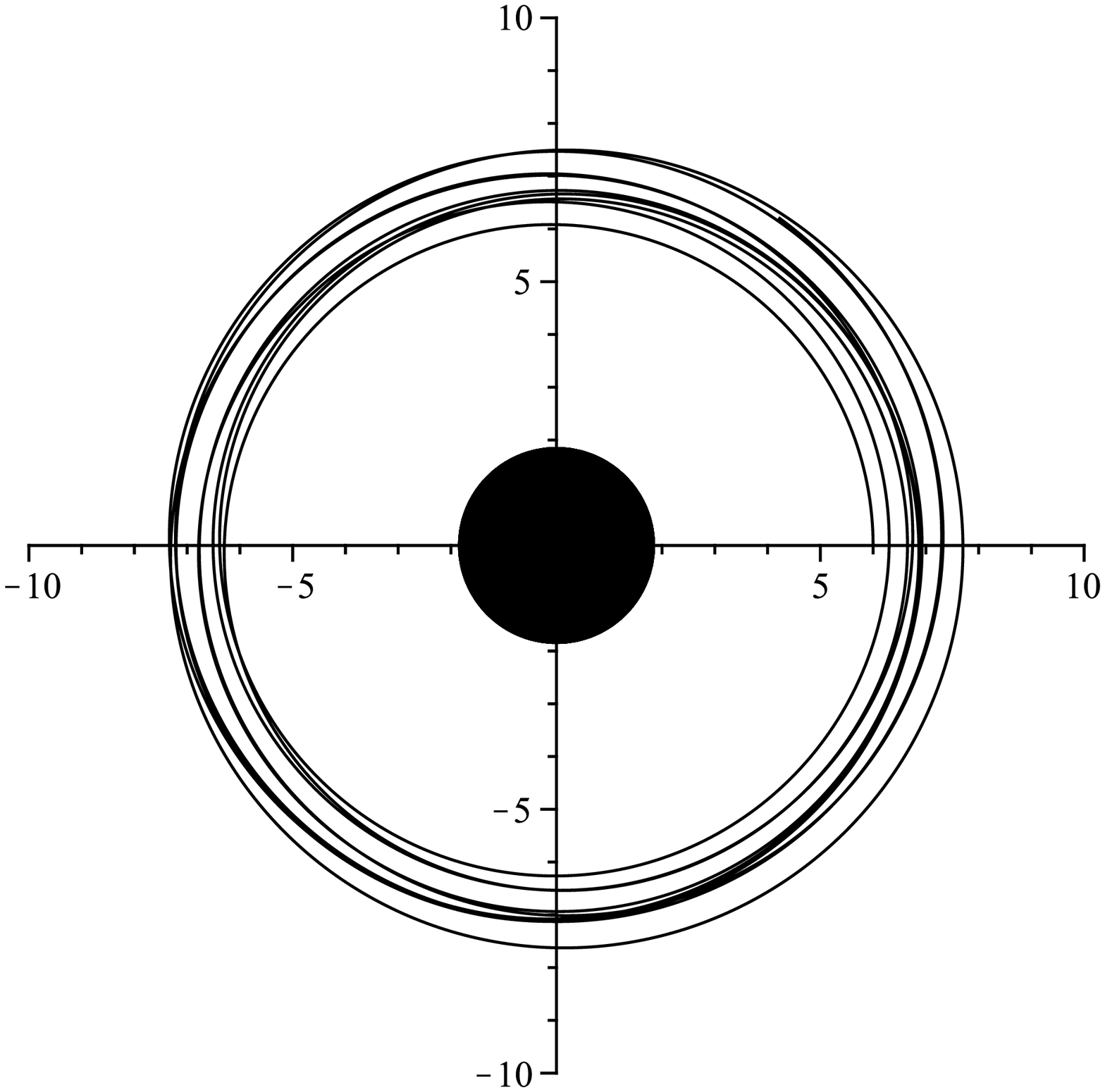}&\quad
\includegraphics[scale=0.3]{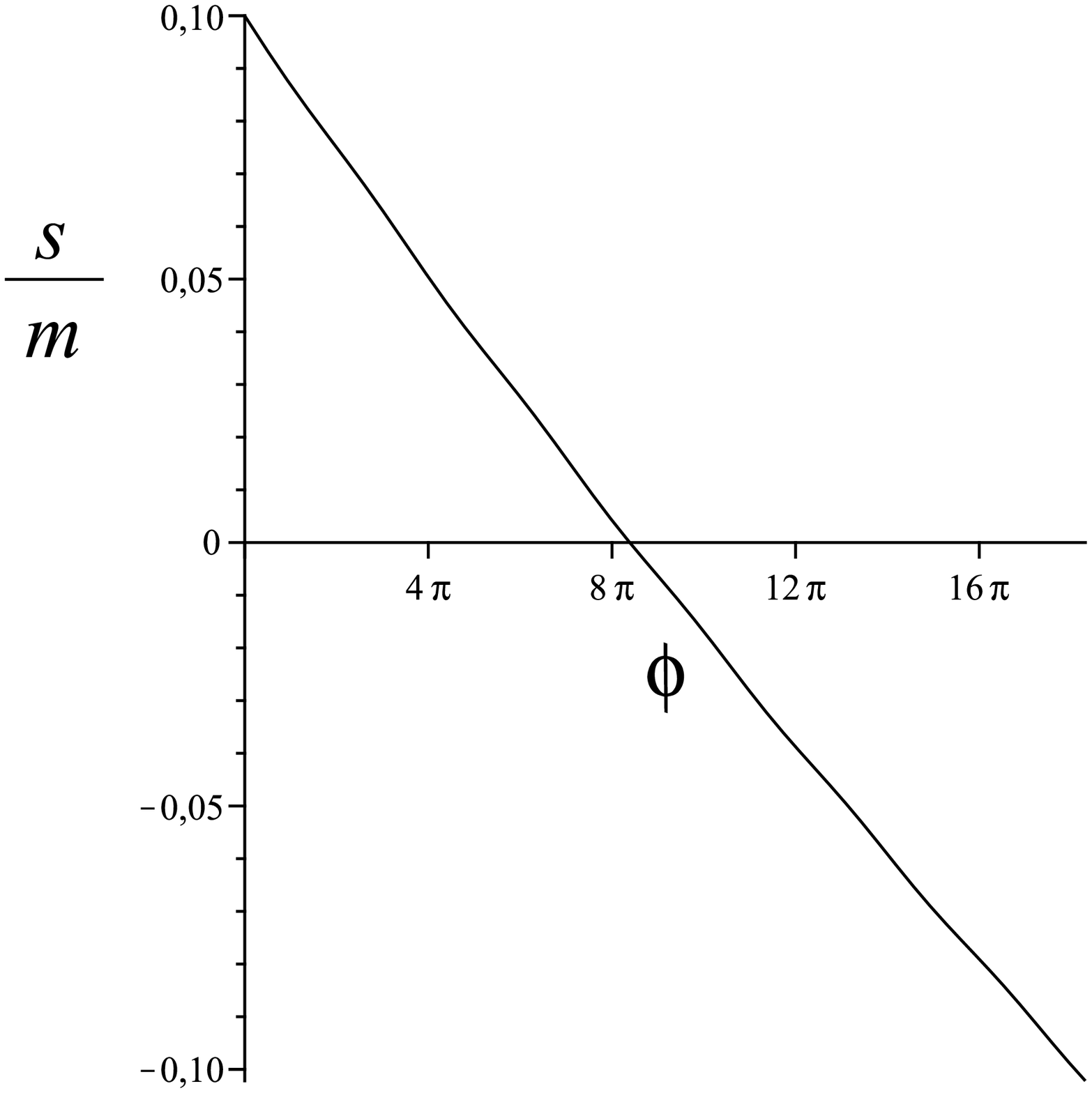}\\[.4cm]
\quad\mbox{(a)}\quad &\quad \mbox{(b)}
\end{array}$\\
\end{center}
\caption{The same as in Fig. \ref{fig:4}, but for a greater initial value of the signed spin magnitude $s(0)=0.1$.
The spin-flip-like effect induced by the non-diagonal components of the quadrupole tensor is here more evident than above.
}
\label{fig:5}
\end{figure}

The spiraling behavior mostly depends on the presence of non-diagonal quadrupole components. In fact, the same choice of quadrupole parameters as in Fig. \ref{fig:2}, except for non-diagonal components which have been switched to zero, leads to oscillations about the initial circular geodesic with very small amplitude (see Fig. \ref{fig:3}).
Reversing the sign of the off-diagonal quadrupole components is responsible for the occurrence of an outward spiraling behavior as well as subtraction of body's spin angular momentum (see Fig. \ref{fig:4}).
This is even more transparent from Fig. \ref{fig:5}, where the outcome of the evolution results in a change of sign (flip) of the spin of the body.
Such an interesting feature was already discussed in the context of interaction between gravitational waves and extended bodies with application to the observed phenomenology of glitches in pulsars \cite{quadrupgw}.
Actually, more properly one should refer to this effect as a \lq\lq spin-flip-like effect,'' because the signed spin magnitude starts from a given positive value, then continuously decreases to zero, switches its sign and monotonically increases with negative values, the orientation of the spin vector being fixed.
Spin-flip, instead, usually implies a sudden change of the direction of the spin vector, its magnitude remaining constant.

\begin{figure} 
\typeout{*** EPS figure 6}
\begin{center}
\includegraphics[scale=0.35]{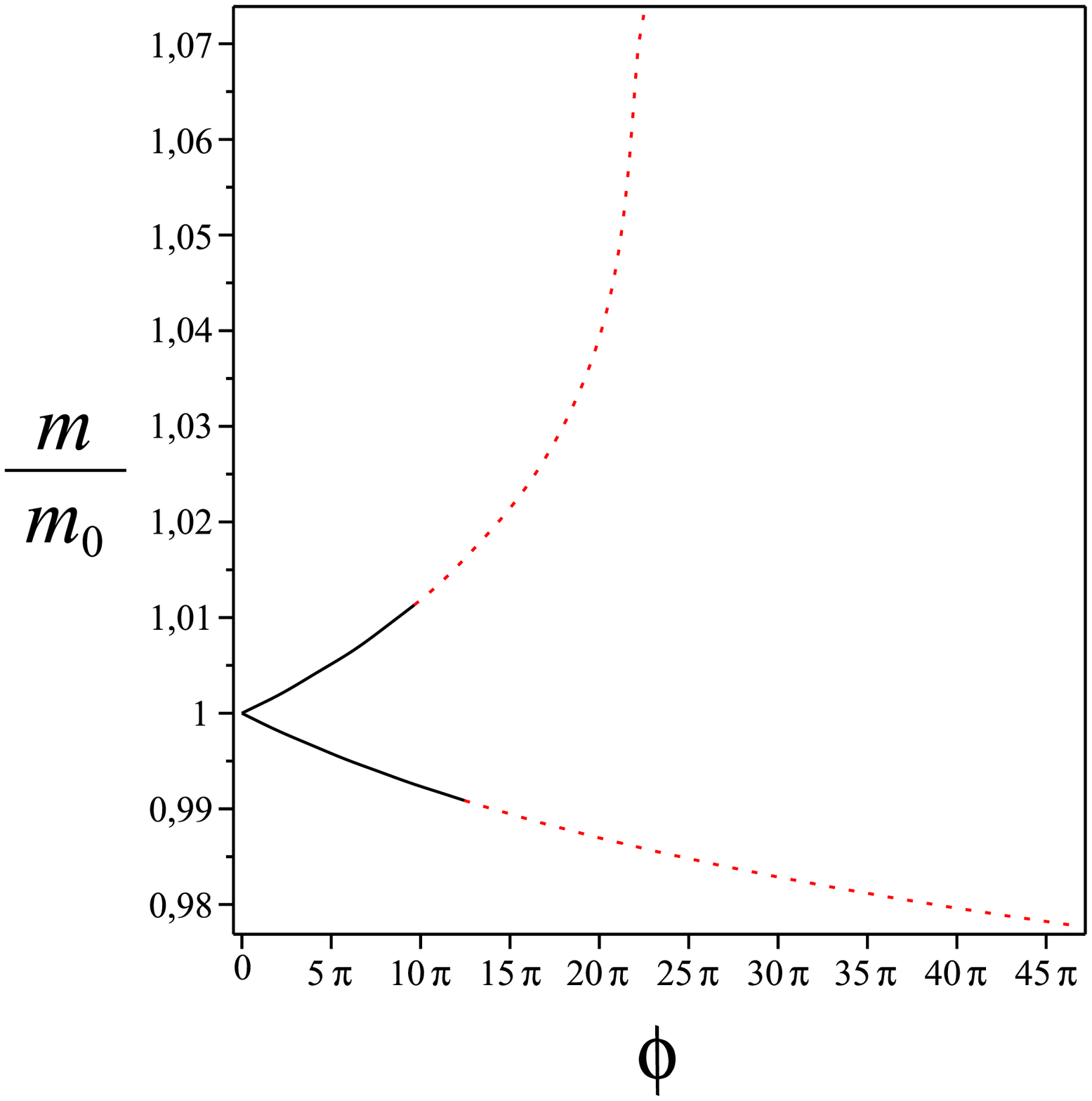}
\end{center}
\caption{The behavior of the particle's dynamical mass $m$ normalized with its initial value $m_0$ is shown for the orbits of Figs. \ref{fig:2} (spiraling in) and \ref{fig:4} (spiraling out).
In the former case the radius of the orbit decreases, whereas both spin and dynamical mass increase (upper curve).
In the latter case one has instead just the opposite (lower curve). 
}
\label{fig:6}
\end{figure}

The features of inward/outward spiraling of the orbits are in agreement with the qualitative discussion of the energy and angular momentum conservation laws at the end of the previous section.
For completeness, we also show in Fig. \ref{fig:6} the evolution of the dynamical mass of the body in the two situations already discussed in Figs. \ref{fig:2} (spiraling in) and \ref{fig:4} (spiraling out). 
In the latter case the particle's mass slightly decreases after each revolution.
In fact, for the chosen set of parameters the ratio $m/m_0$ between the mass $m$ and its initial value $m_0$ passes from 1 to 0.98 in the considered range of $\phi$.
In contrast, if the particle spirals inward as in Fig. \ref{fig:2} the ratio $m/m_0$ slightly increases during the inspiral, the increase becoming more and more enhanced as the horizon is approached (it passes from 1 to 1.07 for the chosen set of parameters).


\section{Concluding remarks}

It has been recently shown that the motion of an extended body on the equatorial plane of a stationary axisymmetric spacetime endowed with reflection symmetry can be described in terms of an effective quadrupole tensor with only 5 nonvanishing components \cite{quadrupkerr1}.
This is due to the combined effect of symmetries of the spacetime and those underlying the MPD model equations.
The quadrupole tensor is thus completely specified by two symmetric and trace-free spatial tensors, namely the mass quadrupole (electric) and the current quadrupole (magnetic) tensors. 
Such a problem has been addressed in the literature in special situations only.
For instance, one can either make certain simplifying assumptions on both the body's structure and dynamics \cite{bfgoschw,bfgokerr} or take into account only spin-induced quadrupole effects, i.e., with a quadrupole tensor proportional to the trace-free part of the square of the spin tensor (see, e.g., Refs. \cite{steinhoff2,hinderer} and references therein).
Therefore, the latter approach concerns only electric-type quadrupole contributions (and of special kind).

We have studied here the role of all effective components of the quadrupole tensor of both kinds on the dynamics of quadrupolar bodies in the equatorial plane of a Kerr spacetime by numerically integrating the full nonlinear set of MPD equations for different spin and quadrupole structure of the body.
A timelike spatially circular geodesic has been chosen as the reference trajectory, i.e., initial conditions have been fixed so that the tangent vector to the orbit of the extended body is initially tangent to the 4-velocity of a circular geodesic.
By construction, the fate of the orbit as described by the MPD model is to remain as close as possible to the reference geodesic, the deviations being confined to a small region around it.
In absence of quadrupole these deviations are actually pure oscillations with amplitude determined by the magnitude of the spin vector, which remains constant along the path just as the mass of the body \cite{mashsingh,spin_dev_schw,spin_dev_kerr}.
A nonzero quadrupole introduces secular effects which enhance deviations, as a general feature.
Furthermore, both the mass and the magnitude of the spin vector are no longer constant.
However, if the quadrupole tensor is represented by its electric (diagonal) part only, the trajectory oscillates, filling a nearly circular corona around the geodesic path, similarly to the case of a purely spinning particle  \cite{quadrupschw,quadrupkerr1}.
As a novel result, the numerical analysis of MPD equations performed here has shown that the non-diagonal (magnetic) components play an important role.
In fact, they are responsible for the occurrence of a spiraling behavior together with a significant increase of the signed magnitude of the spin vector. 
Therefore, in general the limit of validity of the MPD model is reached somewhere during the evolution, the smallness condition of the spin length with respect to the natural background length scale being violated.
In that case backreaction effects should be taken into account. 
Nevertheless, there also exist intermediate situations in which pushing inward and outward compensate each other (instead of competing) for long times, the spin magnitude varying slowly. The body is thus allowed to complete many revolutions around the central object before the MPD regime being lost.
As an interesting feature the spin of the body, which is assumed to be aligned with the spacetime rotation axis, may change its sign during the evolution, leading to a spin-flip-like effect which can be eventually observed.
This effect recalls the phenomenon of pulsar glitches, i.e., a sudden increase in the pulsar rotation frequency, often accompanied by an increase in slow-down rate.
Currently, only multiple glitches of the Crab and Vela pulsars have been observed and studied extensively, even if the physical mechanism triggering glitches is not well understood yet (see, e.g., Refs. \cite{anderson,alpar,zou} and references therein).
The MPD model for an extended body endowed with multipolar structure can be used to describe real astrophysical objects like pulsars or neutron stars. However, in this case the nuclear and hydrodynamical processes occurring in their interior cannot be neglected and should be taken into due account. This is beyond the scope of the present work and is left for future investigations.

\section*{Acknowledgments}
We acknowledge useful discussions with Profs. T. Damour and G. Faye at the beginning of the present project.

\appendix

\section{ZAMO relevant quantities}

We list below the non-vanishing components of the electric and magnetic parts of the Riemann tensor as well as the relevant kinematical quantities as measured by ZAMOs evaluated on the equatorial plane.

The radial components of the acceleration and expansion vectors are given by
\beq\fl\quad
a(n)^{\hat r}=\frac{M}{r^2\sqrt{\Delta}}\frac{(r^2+a^2)^2-4a^2Mr}{r^3+a^2r+2a^2M}\,,\quad
\theta_\phi(n)^{\hat r}=-\frac{aM(3r^2+a^2)}{r^2(r^3+a^2r+2a^2M)}\,,
\eeq
whereas the Lie relative curvature is 
\beq
k_{\rm (lie)}=-\frac{(r^3-a^2M)\sqrt{\Delta}}{r^2(r^3+a^2r+2a^2M)}\,.
\eeq
Finally, the nontrivial components of the electric and magnetic parts of the Riemann tensor with respect to ZAMOs are given by
\begin{eqnarray}
\label{E_H}\fl
E_{\hat r \hat r}&=& -\frac{M(2 r^4+5 r^2 a^2-2 a^2 M r+3 a^4)}{r^4 (r^3+a^2 r+2 a^2 M)}\,, \quad
E_{\hat \theta \hat \theta}=-E_{\hat \phi \hat \phi}- E_{\hat r \hat r}\,,\quad
E_{\hat \phi \hat \phi}=\frac{M}{r^3}\,,\nonumber\\
\fl 
H_{\hat r \hat \theta}&=&  -\frac{3 M a (r^2+a^2) \sqrt{\Delta}}{r^4 (r^3+a^2 r+2 a^2 M)}\,.
\end{eqnarray}

In the limit of vanishing rotation parameter ($a\to0$), the previous quantities simplify to
\beq\fl\quad
a(n)^{\hat r}=\frac{M}{Nr^2}\,,\quad
\theta_\phi(n)^{\hat r}=0\,,\quad
k_{\rm (lie)}=-\frac{N}{r}\,, \quad
E_{\hat r \hat r}=-\frac{2M}{r^3}\,,\quad
H_{\hat r \hat \theta}=0\,,
\eeq
where $N=\sqrt{1-{2M}/{r}}$.

\section{Frame components of both spin and quadrupole terms}

We list below the explicit expressions of the components of both spin and quadrupole terms with respect to the frame adapted to $u$.

The spin force is given by Eq. (\ref{fspinframeu}) with 
\begin{eqnarray}\fl\qquad
\label{fspinframeu2}
F_{\rm (spin)}^1&=&s\gamma\gamma_u\left\{
\nu_u\cos2\alpha_u E_{\hat r\hat r}
+[\nu\cos(\alpha_u-\alpha)+\nu_u\cos^2\alpha_u]E_{\hat \theta\hat \theta}\right.\nonumber\\
\fl\qquad
&&\left.
+\sin\alpha_u[1+\nu\nu_u\cos(\alpha_u-\alpha)]H_{\hat r\hat \theta}
\right\}\,,\nonumber\\
\fl\qquad
F_{\rm (spin)}^2&=&-s\gamma\gamma_u^2\left\{
\nu_u[\sin2\alpha_u-\nu\nu_u\sin(\alpha_u+\alpha)]E_{\hat r\hat r}\right.\nonumber\\
\fl\qquad
&&\left.
+[\nu\sin(\alpha_u-\alpha)+\nu_u\cos\alpha_u(\sin\alpha_u-\nu\nu_u\sin\alpha)]E_{\hat \theta\hat \theta}\right.\nonumber\\
\fl\qquad
&&\left.
-[\cos\alpha_u(1+\nu\nu_u\cos(\alpha_u-\alpha))-2\nu\nu_u\cos\alpha)]H_{\hat r\hat \theta}
\right\}\,,
\end{eqnarray}
the remaining component $F_{\rm (spin)}^0$ following from the condition $F_{\rm (spin)}\cdot U=0$, i.e., 
\begin{eqnarray}\fl
\gamma_u(1-\nu\nu_u\cos(\alpha_u-\alpha))F_{\rm (spin)}^0&=&F_{\rm (spin)}^1\nu\sin(\alpha_u-\alpha)\nonumber\\
\fl
&&
+F_{\rm (spin)}^2\gamma_u(-\nu_u+\nu\cos(\alpha_u-\alpha))\,.
\end{eqnarray}
The spin quantity $D_{\rm (spin)}$ is instead given by Eq. (\ref{dspinframeu}) with 
\beq
\label{dspinframeu3}
{\mathcal E}(u)_{\rm (spin)}=m\gamma\left[\nu\sin(\alpha_u-\alpha)\omega^{1}
+\gamma_u(\nu\cos(\alpha_u-\alpha)-\nu_u)\omega^{2}\right]\,.
\eeq

Concerning the quadrupole terms, the force is given by Eq. (\ref{fquadframeu}) with
\begin{eqnarray}\fl
\label{fquadframeu3}
F_{\rm (quad)}^1&=&
\frac{1}{3}Y_1[b_1\sin3\alpha_u-(b_1-2b_2)\sin\alpha_u]
+\frac{2}{3}Y_2(b_5+b_4\cos2\alpha_u)\nonumber\\
\fl
&&
+\frac{2}{3}\gamma_uY_3[b_1\cos3\alpha_u-(b_1-b_3)\cos\alpha_u]
-\frac{2}{3}\gamma_uY_4b_4\sin2\alpha_u\nonumber\\
\fl
&&
+\frac{2}{3}(2b_2-b_3)X(u)_{22}\sin\alpha_u
\,,\nonumber\\
\fl
F_{\rm (quad)}^2&=&
\frac{1}{3}\gamma_uY_1[b_1\cos3\alpha_u+(4a_2+b_1+2b_2-2b_3)\cos\alpha_u]
+\frac{2}{3}\gamma_uY_2(a_1-b_4)\sin2\alpha_u\nonumber\\
\fl
&&
-\frac{2}{3}\gamma_u^2Y_3[b_1\sin3\alpha_u+(2a_2+b_1-b_3)\sin\alpha_u]
+\frac{2}{3}\gamma_u^2Y_4[b_5+(2a_1-b_4)\cos2\alpha_u]\nonumber\\
\fl
&&
-\frac{2}{3}\gamma_uY_5a_1\sin2\alpha_u
-\frac{8}{3}a_1W(u)_{23}\cos2\alpha_u
+\frac{4}{3}a_2X(u)_{12}\sin\alpha_u\nonumber\\
\fl
&&
+\frac{2}{3}\gamma_u[-2a_2Y_6+(6a_2+2b_2-b_3)X(u)_{22}]\cos\alpha_u
\,,
\end{eqnarray}
with
\begin{eqnarray}\fl\qquad
a_1&=&(E_{\hat \theta\hat \theta}+2E_{\hat r\hat r})\theta_{\hat\phi}(n)^{\hat r}
+H_{\hat r\hat \theta}a(n)^{\hat r}
\,,\nonumber\\
\fl\qquad
a_2&=&(2E_{\hat \theta\hat \theta}+E_{\hat r\hat r})a(n)^{\hat r}
-H_{\hat r\hat \theta}\theta_{\hat\phi}(n)^{\hat r}
\,,\nonumber\\
\fl\qquad
b_1&=&-2(E_{\hat \theta\hat \theta}+2E_{\hat r\hat r})k_{\rm (lie)}
+(2E_{\hat \theta\hat \theta}+E_{\hat r\hat r})a(n)^{\hat r}
-4H_{\hat r\hat \theta}\theta_{\hat\phi}(n)^{\hat r}
+\frac12\partial_{\hat r}E_{\hat \theta\hat \theta}
\,,\nonumber\\
\fl\qquad
b_2&=&-(E_{\hat \theta\hat \theta}+2E_{\hat r\hat r})k_{\rm (lie)}
+2H_{\hat r\hat \theta}\theta_{\hat\phi}(n)^{\hat r}
-\frac32\partial_{\hat r}E_{\hat \theta\hat \theta}
\,,\nonumber\\
\fl\qquad
b_3&=&-2(E_{\hat \theta\hat \theta}+2E_{\hat r\hat r})k_{\rm (lie)}
-2H_{\hat r\hat \theta}\theta_{\hat\phi}(n)^{\hat r}
\,,\nonumber\\
\fl\qquad
b_4&=&H_{\hat r\hat \theta}k_{\rm (lie)}
-3E_{\hat \theta\hat \theta}\theta_{\hat\phi}(n)^{\hat r}
+\partial_{\hat r}H_{\hat r\hat \theta}
\,,\nonumber\\
\fl\qquad
b_5&=&H_{\hat r\hat \theta}k_{\rm (lie)}
-(E_{\hat \theta\hat \theta}+2E_{\hat r\hat r})\theta_{\hat\phi}(n)^{\hat r}
-\partial_{\hat r}H_{\hat r\hat \theta}
\,,
\end{eqnarray}
and 
\begin{eqnarray}\fl\qquad
\label{Yadefs}
Y_1&=&(4\nu_u W(u)_{13}+2X(u)_{11}+X(u)_{22})\gamma_u^2-X(u)_{11}-2X(u)_{22}
\,,\nonumber\\
\fl\qquad
Y_2&=&[4W(u)_{13}+\nu_u(2X(u)_{11}+X(u)_{22})]\gamma_u^2-2W(u)_{13}
\,,\nonumber\\
\fl\qquad
Y_3&=&X(u)_{12}+2\nu_u W(u)_{23}
\,,\nonumber\\
\fl\qquad
Y_4&=&\nu_u X(u)_{12}+2W(u)_{23}
\,,\nonumber\\
\fl\qquad
Y_5&=&2W(u)_{13}+\nu_u(X(u)_{11}+2X(u)_{22})
\,,\nonumber\\
\fl\qquad
Y_6&=&2\nu_u W(u)_{13}+X(u)_{11}+2X(u)_{22}
\,.
\end{eqnarray}
The remaining component $F_{\rm (quad)}^0$ can be obtained from the vanishing of the coordinate component $F_{{\rm (quad)}\,t}=0$, which implies
\begin{eqnarray}\fl\qquad
0&=&\gamma_u(F_{\rm (quad)}^0+\nu_u F_{\rm (quad)}^2)\nonumber\\
\fl\qquad
&&+\frac{\sqrt{g_{\phi\phi}}N^{\phi}}{N}\left[
F_{\rm (quad)}^1\cos\alpha_u
-\gamma_u\sin\alpha_u(\nu_u F_{\rm (quad)}^0+F_{\rm (quad)}^2)
\right]\,.
\end{eqnarray}
Finally, the torque term is given by Eqs. (\ref{dquadframeu})--(\ref{dquadframeu2}) with components
\begin{eqnarray}\fl
\label{quadtorquecompts}
{\mathcal E}(u)_{\rm (quad)}{}_1 &=&
\frac{2}{3}(2E_{\hat r\hat r}+E_{\hat \theta\hat \theta})[(Y_4\gamma_u^2-4W(u)_{23})\cos2\alpha_u
-Y_5\gamma_u\sin2\alpha_u]\nonumber\\
\fl
&&
+\frac{4}{3}H_{\hat r\hat \theta}[Y_6\gamma_u\cos\alpha_u
+(2Y_3\gamma_u^2-X(u)_{12})\sin\alpha_u]
+2\gamma_u^2E_{\hat \theta\hat \theta}Y_4
\,,\nonumber\\
\fl
{\mathcal E}(u)_{\rm (quad)}{}_2 &=&
-\frac{2}{3}(2E_{\hat r\hat r}+E_{\hat \theta\hat \theta})[Y_2\cos2\alpha_u
-Y_4\gamma_u\sin2\alpha_u]\nonumber\\
\fl
&&
-\frac{4}{3}H_{\hat r\hat \theta}[Y_3\gamma_u\cos\alpha_u
+(2Y_1+3X(u)_{22})\sin\alpha_u]
-2E_{\hat \theta\hat \theta}Y_2
\,,\nonumber\\
\fl
{\mathcal B}(u)_{\rm (quad)}{}_3 &=&
\frac{2}{3}(2E_{\hat r\hat r}+E_{\hat \theta\hat \theta})[(Y_3\gamma_u^2+X(u)_{12})\cos2\alpha_u
+(Y_6-3X(u)_{22})\gamma_u\sin2\alpha_u]\nonumber\\
\fl
&&
-\frac{4}{3}H_{\hat r\hat \theta}[(Y_5-3\nu_uX(u)_{22})\gamma_u\cos\alpha_u
-2(Y_4\gamma_u^2-W(u)_{23})\sin\alpha_u]\nonumber\\
\fl
&&
+2E_{\hat \theta\hat \theta}(Y_3\gamma_u^2-X(u)_{12})
\,.
\end{eqnarray}

\subsection{The Schwarzschild limit}

We list below the corresponding expressions of both spin and quadrupole terms in the limit of vanishing spacetime rotation.

The spin force (\ref{fspinframeu}) becomes
\begin{eqnarray}\fl\quad
F_{\rm (spin)}&=&\frac{M}{r^3}\gamma\gamma_u s\left\{
3\gamma_u\nu_u\sin\alpha_u(\nu\cos\alpha-\nu_u\cos\alpha_u)u\right.\nonumber\\
\fl\quad
&&\left.
+\frac12[\nu_u(1-3\cos2\alpha_u)+2\nu\cos(\alpha_u-\alpha)]e_1\right.\nonumber\\
\fl\quad
&&\left.
-\frac12\gamma_u[(2+\nu_u^2)\nu\sin(\alpha_u-\alpha)+3\nu_u(\nu\nu_u\sin(\alpha_u+\alpha)-\sin2\alpha_u)]e_2
\right\}\,,
\end{eqnarray}
whereas the spin quantity $D_{\rm (spin)}$ is still given by Eq. (\ref{dspinframeu}) with components (\ref{dspinframeu3}).

The quadrupole force (\ref{fquadframeu}) becomes
\beq
F_{\rm (quad)}=F_{\rm (quad)}^1e_1+F_{\rm (quad)}^2(-\nu_u u+e_2)\,,
\eeq
where $\gamma_u(-\nu_u u+e_2)$ represents a unitary and spacelike vector orthogonal to $n$ and 
\begin{eqnarray}\fl\quad
F_{\rm (quad)}^1&=&
-\frac{2MN}{r^4}
[(1-5\sin^2\alpha_u)(Y_1\sin\alpha_u+2Y_3\gamma_u\cos\alpha_u)\nonumber\\
\fl\quad
&&+(Y_1-3X(u)_{22})\sin\alpha_u]
\,,\nonumber\\
\fl\quad
F_{\rm (quad)}^2&=&
\frac{MN}{r^4}\gamma_u
[5\sin2\alpha_u(Y_1\sin\alpha_u+2Y_3\gamma_u\cos\alpha_u)\nonumber\\
\fl\quad
&&-2(2Y_3\gamma_u\sin\alpha_u-3X(u)_{22}\cos\alpha_u)]
\,,
\end{eqnarray}
where the lapse function is $N=\sqrt{1-{2M}/{r}}$ and the coefficients $Y_a$ are still given by Eq. (\ref{Yadefs}).
Finally, the torque term is given by Eqs. (\ref{dquadframeu})--(\ref{dquadframeu2}) with components
\begin{eqnarray}\fl\qquad
{\mathcal E}(u)_{\rm (quad)}{}_1 &=&
\frac{4M}{r^3}[\gamma_u\sin\alpha_u(Y_4\gamma_u\sin\alpha_u+Y_5\cos\alpha_u)+2W(u)_{23}\cos2\alpha_u]
\,,\nonumber\\
\fl\qquad
{\mathcal E}(u)_{\rm (quad)}{}_2 &=&
-\frac{4M}{r^3}\sin\alpha_u(Y_2\sin\alpha_u+Y_4\gamma_u\cos\alpha_u)
\,,\nonumber\\
\fl\qquad
{\mathcal B}(u)_{\rm (quad)}{}_3 &=&
\frac{4M}{r^3}\{\gamma_u\sin\alpha_u[Y_3\gamma_u\sin\alpha_u-(Y_6-3X(u)_{22})\cos\alpha_u]
\nonumber\\
\fl\qquad
&&
-X(u)_{12}\cos^2\alpha_u\}
\,.
\end{eqnarray}

\end{document}